\documentclass{biophys-new}
\usepackage[utf8]{inputenc}
\usepackage{graphicx}
\usepackage{xr-hyper}
\usepackage[colorlinks,allcolors=cyan!70!black]{hyperref}
\usepackage{lipsum}
\usepackage{verbatim}
\usepackage[colorinlistoftodos,prependcaption,textsize=tiny]{todonotes}
\usepackage[labelfont=bf]{caption}
\usepackage{amsfonts, amsmath}

\makeatletter

\newcommand*{\addFileDependency}[1]{
\typeout{(#1)}
\@addtofilelist{#1}
\IfFileExists{#1}{}{\typeout{No file #1.}}
}\makeatother

\newcommand*{\myexternaldocument}[1]{%
\externaldocument{#1}%
\addFileDependency{#1.tex}%
\addFileDependency{#1.aux}%
}

\myexternaldocument{supplementalv3}

\title{Bind-and-bend model for DNA looping}
\runningtitle{Bind-and-bend model}

\author[1]{Michael L. Liu}
\author[1]{Daniel W. Oo}
\author[1]{Ryan B. McMillan}
\author[1,*]{Ashley R. Carter}
\corrauthor[*]{acarter@amherst.edu}
\affil[1]{Department of Physics, Amherst College, Amherst, MA 01002}
\runningauthor{Liu et al.}

\papertype{Article}

\begin{document}

\begin{frontmatter}

\begin{abstract}

DNA looping is important in DNA condensation and regulation. One method for forming a DNA loop, thought to be used by the condensing agent protamine, is bind-and-bend. In bind-and-bend, molecules bind all along the DNA, each creating a bend in the DNA. Eventually, enough bending leads to the formation of a loop. Here, we adapt theory for DNA bending by cations to create a simple bind-and-bend model. To test the model, we simulate bending and looping by the condensing agent protamine and compare the output of the simulation to experimental data. The model captures several interesting features of the data including: the curvature of the DNA due to both protamine-induced bending and thermal fluctuations, the small circumference of the loops (200-300 bp), the bias in the location where the loop forms, and the emergence of multi-looped flower structures. The model leads to insight into where protamine binds, how it bends DNA, and how it creates one or more DNA loops. More broadly, the model could be useful in understanding the compaction of nucleic acids or polyelectrolytes.

\end{abstract}

\begin{sigstatement}
In biology, nucleic acid compaction is important in forming chromosomes during mitosis, in packaging genetic material into viruses, in creating a hydrodynamic sperm during spermatogenesis, and in regulating gene expression. In nanoengineering, nucleic acid compaction is useful in designing biosensors or drug delivery vehicles. In materials research, compaction of polyelectrolytes by counter-ions is important in the formation of coacervates. Here we look at one type of compaction, the compaction of double-stranded DNA into loops by protamine. Specifically, we develop a model for protamine to bind-and-bend the DNA into a loop. This model gives insight into the physics of how positively charged molecules work together to bind and bend a negatively charged polymer, which could be a nucleic acid or polyelectrolyte.

\end{sigstatement}

\end{frontmatter}

\twocolumn

\section*{Introduction}

DNA looping is important in DNA regulation \cite{cournac2013DNA,schleif1992DNA,matthews1992DNA, saiz2006DNA}, organization \citep{ruiten_smc_2018, hassler_towards_2018, yuen_taking_2018, terakawa_condensin_2017, krepel_deciphering_2018}, and condensation \citep{rimsky_structure_2004, balhorn_protamine_2007, dame_bacterial_2006, dame_dna_2005}. In prokaryotes and eukaryotes, DNA looping regulates gene expression by increasing the local concentration of regulatory proteins near the promoter and facilitating interactions with RNA polymerase \cite{cournac2013DNA,schleif1992DNA,matthews1992DNA, saiz2006DNA}. During mitosis, DNA looping is important in the assembly of chromosomes \cite{ruiten_smc_2018, hassler_towards_2018, yuen_taking_2018, terakawa_condensin_2017, krepel_deciphering_2018}. In sperm, DNA looping is one of the first steps in a process that condenses the entire genome in minutes \cite{hud_toroidal_2005, balhorn_protamine_2007, pogany_dna_1981, vilfan_formation_2004}. In bioengineering, DNA loops have been used in schemes for drug delivery \cite{bastings2018modulation}.

There are several mechanisms for DNA loop formation. One mechanism, employed by bacterial transcription factors like \emph{lac} repressor \cite{cournac2013DNA, becker2013mechanism, wong2008interconvertible, friedman1995crystal}, is to bind one location on the DNA, wait for spontaneous thermal fluctuations to fold the DNA over, and then bind a second location on the folded DNA to form a loop. Another mechanism, used by motor proteins like condensin, is to use the energy of ATP hydrolysis to extrude a loop \cite{ganji2018real}. Finally, a third mechanism, used by the condensing agent protamine \cite{ukogu2020protamine}, is bind-and-bend. In bind-and-bend, multiple molecules bind and bend the DNA, leading to the formation of a loop. 

It is not known whether other condensing agents besides protamine use bind-and-bend to loop DNA. Condensing agents are positively charged molecules that bind the negatively charged DNA and condense it into a toroid \cite{hud_toroidal_2005}, which is a series of hexagonally packed, stacked loops \cite{hud2001cryoelectron}. Examples of condensing agents include multivalent cations like cobalt hexaammine (III) \cite{schnell1998insertion, bloomfield_condensation_1991}, polyamines like spermine \cite{bloomfield_condensation_1991, leforestier_structure_2009, leforestier2011protein, takahashi1997discrete} and spermidine \cite{fang1998early, murayama2003elastic, marx1983evidence, takahashi1997discrete}, and arginine-rich proteins like protamine \cite{ukogu2020protamine, allen1997afm, balhorn_protamine_2007}. The toroids formed by condensing agents are at almost crystalline packing levels \cite{hud_toroidal_2005, teif2011condensed, sotolongo_ability_2003}, have an inner diameter of 30-50 nm, an outer diameter of 80-100 nm, and contain $\sim50$ kbp of DNA \cite{teif2011condensed, hud_toroidal_2005, sotolongo_ability_2003, yoshikawa1996nucleation, allen1997afm}. There are several models for toroid formation \cite{hud_toroidal_2005, mcmillan_dna_nodate, fang1998early, sung2011condensation, golan1999dna, ou2005langevin}, but all agree that condensing agents act by binding all along the DNA and neutralizing the negatively charged DNA backbone \cite{hud_toroidal_2005}. 

Charge neutralization by condensing agents could cause two effects: DNA bending \cite{bloomfield_condensation_1991, rouzina1998dna, bloomfield1997dna, ukogu2020protamine, mcmillan_dna_nodate, mukherjee2021protamine} or DNA-DNA interactions \cite{hud_toroidal_2005, van2010visualizing}. DNA bending by small cations is thought to be due to an electrostatic effect that causes the negative charges on the DNA backbone to be attracted to the positively charged condensing agent bound in the major groove of the DNA \cite{rouzina1998dna}. This attraction closes the groove and is thought to create a bend in the DNA of $20^{\circ}\text{-}40^{\circ}$. DNA-DNA interactions are interactions between the DNA strands that facilitate DNA packing into the toroid \cite{hud_toroidal_2005, van2010visualizing} and also stabilize the DNA crossover point at the close of the loop. These DNA-DNA interactions could be due to Van der Waals interactions between the neutral DNA strands \cite{manning_counterion_2007}, electrostatic interactions between one condensing agent molecule and two adjacent DNA strands \cite{mukherjee2021protamine}, or covalent interactions (disulfide bridges) between two condensing agent molecules bound on adjacent DNA strands \cite{balhorn_protamine_2007, oliva_vertebrate_1991}. These two effects, DNA bending and DNA-DNA interactions, should work together to create the DNA loop.

Our goal is to better understand how protamines, and possibly other condensing agents, use bind-and-bend to loop DNA. In particular, we are interested in the physics involved in four different areas: the binding of DNA, the bending of DNA into a loop, the stability of the DNA-DNA interaction, and the formation of multiple loops. 

To make DNA looping predictions, we adapt the theory on DNA bending by small cations \cite{rouzina1998dna} to create a simple simulation of bind-and-bend in 2D. Specifically, we represent the DNA as a planar, piece-wise linear curve. Then, we assume the DNA is a worm-like chain \cite{marko1995stretching} and add spontaneous thermal fluctuations to the curve based on a Boltzmann probability distribution using the elastic energy of DNA \cite{rouzina1998dna}. Next, we represent the condensing agent as a positively charged sphere and use another Boltzmann probability distribution to locate a binding site on the DNA. The total energy in the Boltzmann distribution is given by the eletrostatic potential energy between the condensing agent and the DNA, the entropic cost associated with localizing the condensing agent, and the bending of the DNA \cite{rouzina1998dna}. We position the condensing agent given this Boltzmann distribution and bend the DNA. Finally, we repeat this process for other condensing agent molecules, creating a simulated DNA contour. 

To test the model, we set the simulation parameters to model a DNA-protamine system and compare the simulated DNA contours to our experimental data set \cite{mcmillan_dna_nodate, mcmillan2021dna, ukogu2020protamine}. We find that the model captures the interesting features of the data. The model is able to predict a double peak in the radius of curvature of the DNA, with one peak due to DNA bending by protamine and one peak due to spontaneous thermal fluctuations. The model is also able to predict the small loop circumference (200-300 bp), the spatial distribution of loops, and the formation of DNA flowers. Creation of the model leads to insights about how protamine binds and bends DNA, as well as how DNA might be folded by condensing agents more generally.

\begin{figure}[!htb]
\centering
\includegraphics[width=0.92\linewidth]{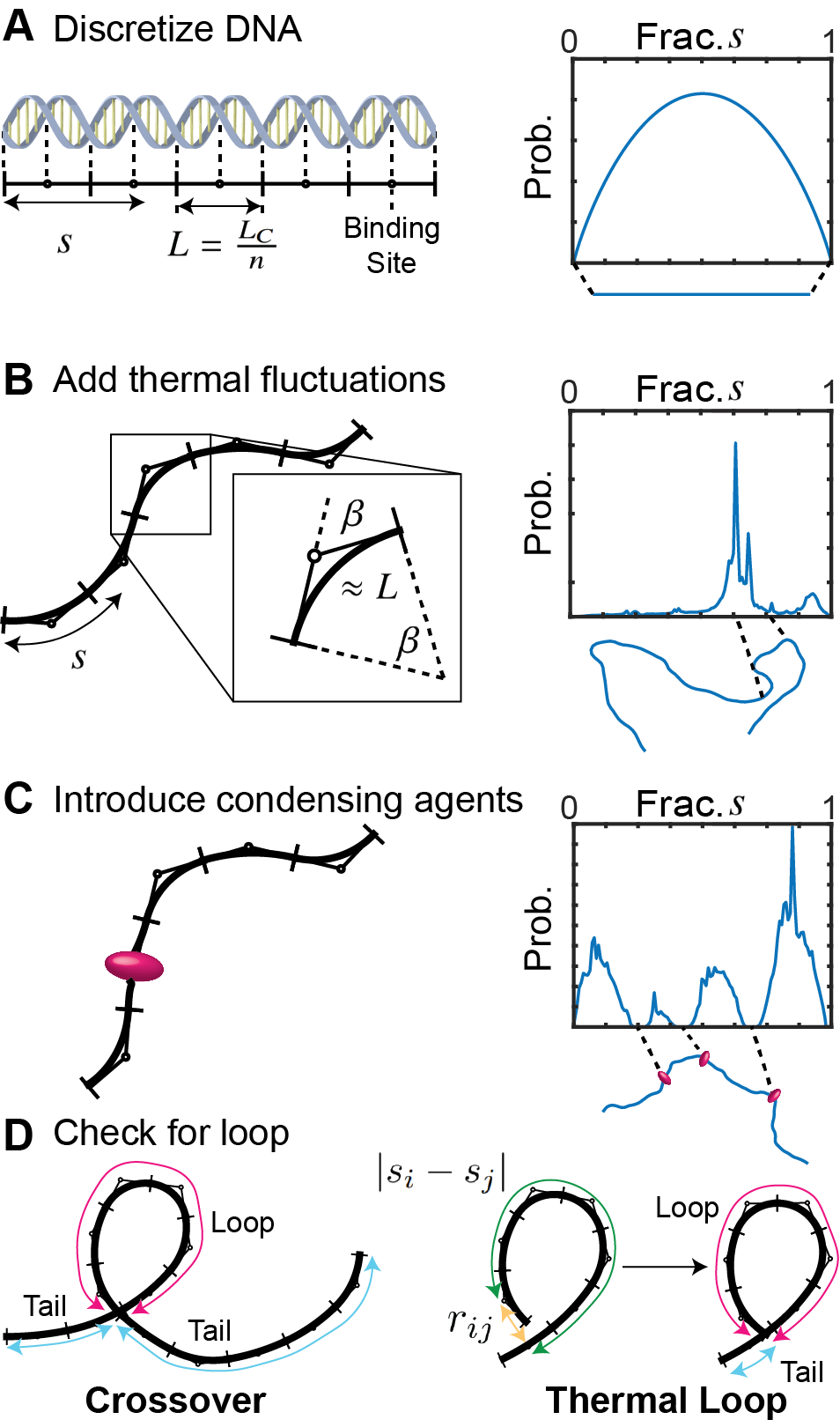}
\caption{\textbf{Method to simulate DNA contours and check for loops.} A) DNA with contour length $L_C$ is discretized into $n$ regions of length $L$. Binding sites are at the midpoints of the regions. Length along the DNA contour is $s$. For straight, 1D DNA, the highest binding probability for condensing agents is at a fractional length $s/L_C$ (Frac. $s$) of 0.5. B) We add thermal fluctuations which bend each region at its midpoint, creating a circular arc with subtended angle $\beta$. For curved, 2D DNA, the highest binding probability is at locations with a high local concentration of DNA. C) Next, condensing agents bind and bend the DNA. Binding probability decreases at the bound location. D) After generating contours, we check for loops. There are two mechanisms of loop formation: ``crossover'' and ``thermal loop closing''. Thermal loop closing requires two distal DNA regions $i,j$ to have planar distance $r_{ij}< k_\text{planar}L_P$ and contour distance $|s_i - s_j|>k_\text{linear} L_P$.}
\label{fig:Methods}
\end{figure}

\section*{Theory}

\subsection*{Discretizing the DNA molecule}

Consider a single molecule of double stranded DNA with contour length $L_C$. We partition the DNA into $n$ jointed regions of length $L = \frac{L_C}{n}$. At the midpoint of each region, we place a binding site (Fig. \ref{fig:Methods}A). For the condensing agent protamine, we set $L = 10$ bp (3.4 nm), since protamine is thought to bind DNA every 10 bp or so \cite{bench_dna_1996}.

\subsection*{Adding spontaneous thermal fluctuations}

To generate the initial state of a DNA molecule, we compute bend angles due to spontaneous thermal fluctuations for all $n$ regions. We compute the distribution of these bend angles using a Boltzmann distribution with an energy given by the elastic potential energy of the DNA. 

We use the elastic potential energy previously used by Rouzina and Bloomfield \cite{rouzina1998dna}. Specifically, we calculate the elastic potential energy by assuming the DNA is in thermal equilibrium with the surrounding solution. In this case, the elastic energy $U_\text{elas}$ is given by the general equation \cite{schellman1974flexibility}
\begin{equation} \label{eq:UElasGeneral}
U_\text{elas} (\beta) = \frac{1}{2} g \beta^2,
\end{equation}

\noindent where $g$ is the bending rigidity and $\beta$ is the bend angle. If we assume that the DNA is a worm-like chain \cite{marko1995stretching}, we can replace the bending rigidity with an expression that depends on the Boltzmann constant $k_B$, the ambient temperature $T$, the length of the DNA region $L$, and the persistence length $L_P$. We also assume that the DNA acts as a hinge and only bends in one direction \cite{schellman1974flexibility}. Under these assumptions, the elastic energy $U_\text{elas}$ is given by
\begin{equation} \label{eq:UElasCircle}
U_{\text{elas}}(\beta) = k_B T \frac{L_P}{4 L} \beta ^2.
\end{equation}

\noindent If the DNA is not a hinge and instead has isotropic bending in both directions, the factor of $4L$ becomes $8L$. 

Second, we apply the Boltzmann distribution to Eq.~\ref{eq:UElasCircle} to obtain the probability density function
\begin{equation} \label{eq:BetaPDF}
p(\beta) = \frac{1}{Z} e^{-\frac{L_P}{4 L} \beta^2}
\end{equation}
for $\beta$. Here, $Z$ is the normalization factor that ensures $p(\beta) \in [0, 1]$. Thus, the spontaneous thermal bend $\beta$ is normally distributed with mean $\mu = 0$ and standard deviation $\sigma = \sqrt{\frac{2 L}{L_P}}$, which for protamine corresponds to about $20^\circ$. We select $\beta_i$, the bend angle at the $i^\text{th}$ binding site, from this distribution.

We then bend each region at its binding site, splitting the region into two equal-length, straight segments, one on either side of the binding site. Specifically, the rightmost segment is rotated by the angle $\beta$ relative to the binding site, creating a circular arc of length $\approx L$ and subtended angle $\beta$ (Fig. \ref{fig:Methods}B).

\subsection*{Introducing condensing agents: binding energy}

Next, we introduce molecules of condensing agent to the DNA and determine the binding sites. First, we find the potential energy $U_{\text{bind}, \, i}$ for a single condensing agent molecule to bind to the $i^\text{th}$ binding site. We assume this potential energy contains the electrostatic interactions between the positively charged condensing agent molecules and the negatively charged DNA backbone, as well as the entropic cost to localize a single condensing agent molecule. 

To estimate the potential energy $U_{\text{bind}, \, i}$, we assume that the energy contribution $U_i$ from the $i^\text{th}$ binding site is of a different form than the energy contribution $U_j$ from the $j^\text{th}$ binding site for all $j \neq i$. That is, we model the condensing agent's local interaction with its own binding site differently from its interaction with other, more distant regions on the DNA. Thus, we have
\begin{equation} \label{eq:SumUiUj}
U_{{\text{bind}, \, i}}=U_i+\sum _{j\neq i} U_j.
\end{equation}
We assume that the energy contributions $U_j$ from the $j \neq i$ regions of DNA are given by Coulomb's Law,
\begin{equation} \label{eq:Uj}
U_j = \frac{1}{4 \pi \epsilon _0 K(r_{i j})} \frac{q_{\text{agent}} q_j}{r_{i j}},
\end{equation}

\noindent where $r_{i j}$ is the planar distance between the $i^\text{th}$ and $j^\text{th}$ binding sites, $q_{\text{agent}}$ is the charge of the condensing agent, $q_j$ is the total charge at the $j^\text{th}$ binding site, and $K(r_{ij})$ is a distance dependent dielectric function. We use the expression
\begin{equation} \label{eq:qj}
q_j=\alpha _j p_\text{agent} q_\text{agent}+p_\text{DNA} q_\text{DNA}
\end{equation}
\noindent to calculate the total charge at the $j^\text{th}$ binding site. The variable $\alpha _j \in \{0, 1\}$ accounts for whether the $j^\text{th}$ binding site is occupied by a molecule of condensing agent. The variable $q_{\text{DNA}}$ is the total negative charge on one region of DNA, given by $\lambda L$, where $\lambda$ is the linear charge density of the DNA. The factors $p_{\text{agent}}\in [0, 1]$ and $p_{\text{DNA}}\in [0, 1]$ account for the effective electrostatic screening of the condensing agent and DNA charges, respectively. Similar to previous theory \cite{rouzina1998dna}, we also use the distant-dependent dielectric function $K(r_{ij})$ \cite{hingerty1985dielectric} that depends on the planar distance $r_{i j}$,
\begin{equation} \label{eq:K}
K(r_{i j}) = D - \frac{D-1}{2} \left(\left(\frac{C r_{i j}}{H}\right)^2+2\frac{ C r_{i j}}{H}+2\right) e^{-\frac{C r_{i j}}{H}}.
\end{equation}

\noindent Here, the variables are the bulk dielectric of water $D= 80$, the constant $C= 2.674$, and the half-saturation length of water $H= 0.75$ nm \cite{rouzina1998dna}.

Next, we need an equation for the energy contribution $U_i$ in Eq.~\ref{eq:SumUiUj}. If the $i^\text{th}$ binding site has an existing bend $\beta$ due to random thermal fluctuations, then the energy contribution $U_i$ has an electrostatic term and an entropic term, 
\begin{equation} \label{eq:Ui}
U_i(r) = U_{\text{elec}}(r) - T \Delta S(r).
\end{equation}

\noindent That is, there is an electrostatic stabilization of the condensing agent by the DNA charge at the $i^\text{th}$ binding site, but there is also an entropic cost to localizing the condensing agent at that site. Both of these terms are dependent on $r$, which is the effective separation between the middle of the condensing agent and the edge of the DNA rod. This separation $r$ can be approximated as a linear function of the DNA bend angle $\beta$,
\begin{equation} \label{eq:r}
r(\beta) = r_0 - \frac{R_\text{DNA}}{2} |\beta|.
\end{equation} 

\noindent Here, $r_0$ is half the width of the groove, and $R_\text{DNA}$ is the radius of the DNA double helix. We set both $r_0$ and $R_\text{DNA}$ to 1 nm, as was done previously \cite{rouzina1998dna}.

We assume that the electrostatic term in Eq.~\ref{eq:Ui} follows Coulomb's Law,
\begin{equation} \label{eq:UElec}
U_\text{elec}(r) = \frac{1}{4 \pi \epsilon _0 K(r)} \frac{p_\text{site} q_\text{agent} q_\text{DNA}}{r},
\end{equation}

\noindent where $p_\text{site} \in [0, 1]$ accounts for the effective electrostatic screening of the charges $q_\text{agent}$ and $q_\text{DNA}$ at the binding site. 

We express the entropic term in Eq.~\ref{eq:Ui} as
\begin{equation} \label{eq:-TDS}
-T \Delta S (r) = \left( \frac{1}{2} \ln \left(-\frac{2 U_{\text{elec}}(r)}{k_B T} \right) + \frac{1}{2} - \ln (q) \right) k_B T,
\end{equation}

\noindent where $q$ is the frequency of localizable condensing agent per binding site \cite{rouzina1998dna}, which we take to be constant. 

Now, substituting expressions for $U_i$ and $U_j$, we may rewrite Eq.~\ref{eq:SumUiUj} as
\begin{multline} \label{eq:UBind}
U_{\text{bind}, \, i} = U_\text{elec}(r) + \left( \frac{1}{2} \ln \left(-\frac{2 U_{\text{elec}}(r)}{k_B T} \right) + \frac{1}{2} - \ln (q) \right) k_B T \\ + \sum _{j\neq i} \frac{1}{4 \pi \epsilon _0 K(r_{i j})} \frac{q_{\text{agent}} q_j}{r_{i j}}.
\end{multline}

\noindent Next, we remove constant terms for simplicity, since we only care about relative changes in the potential energy, rather than the absolute potential energy. In addition, we use Eq.~\ref{eq:r} to rewrite Eq.~\ref{eq:UBind} in terms of $\beta_i$, the bend angle at the $i^\text{th}$ binding site. Thus, we obtain
\begin{multline} \label{eq:UBindSimplified}
U_{\text{bind}, \, i} = U_\text{elec}(\beta_i) + \frac{k_B T}{2} \ln (U_\text{elec}(\beta_i))
\\ + \sum _{j\neq i} \frac{1}{4 \pi \epsilon _0 K(r_{i j})} \frac{q_{\text{agent}} q_j}{r_{i j}}.
\end{multline}

\noindent Note that this equation only depends on the bend angle $\beta_i$ at the $i\text{th}$ region, the planar distance $r_{ij}$ between the $i^\text{th}$ and $j^\text{th}$ region, and the charges $q_\text{agent}$ and $q_j$.

\subsection*{Introducing condensing agents: bending energy}

Thus far, we have addressed the energy change $U_{\text{bind}, \, i}$ as a condensing agent binds to the $i^\text{th}$ site on the DNA. Now, we estimate the energy change $U_{\text{bend}, \, i}$ as a newly bound condensing agent bends the DNA at the $i^\text{th}$ site.

To estimate this bending energy, we use theory developed by Rouzina and Bloomfield \cite{rouzina1998dna}. Specifically, we account for three effects on the energy of the condensing agent-DNA system. First, the bending incurs an elastic cost, given by Eq.~\ref{eq:UElasCircle}, as the DNA stretches beyond its unperturbed state. Second, there is increased electrostatic stabilization of the positively charged condensing agent by the neighboring phosphate groups, given by Eq.~\ref{eq:UElec}, as the effective agent-DNA separation decreases. Third, there is a larger entropic cost of localizing the condensing agent, given by Eq.~\ref{eq:-TDS}, as the DNA groove cavity narrows. Thus, the bending energy is
\begin{gather} 
U_{\text{bend}, \, i} = \Delta U_{\text{elas}} + \Delta U_{\text{elec}} - T \Delta S.
\end{gather}

\noindent Substituting Eqs.~\ref{eq:UElasCircle}, \ref{eq:UElec}, and \ref{eq:-TDS}, removing constant terms, and rewriting gives
\begin{multline}
\label{eq:UBendSimplified}
U_{\text{bend}, \, i} = k_B T \frac{L_P}{4 L} \left(\beta_{\text{final}, \, i}^2-\beta_i^2\right) + U_{\text{elec}}(\beta_{\text{final}, \, i}) - U_{\text{elec}}(\beta_i) \\ + \frac{k_B T}{2} \ln \left(\frac{U_{\text{elec}}(\beta_{\text{final}, \, i})}{U_{\text{elec}}(\beta_i)}\right),
\end{multline}

\noindent where $\beta_{\text{final}, \, i}$ is the final angle at the $i^\text{th}$ binding site after bending of the DNA by condensing agents.

\subsection*{Introducing condensing agents: bend angle}

To calculate $U_{\text{bend}, \, i}$, we need to know the final bend angle $\beta_{\text{final}, \, i}$ due to condensing agent binding. There are several considerations for estimating $\beta_{\text{final}, \, i}$.

The first consideration is to determine if bending can occur in both directions. Here, we set bending in one direction. The experimental data for protamine shows that 80\% of DNA molecules are bent in the same direction \cite{ukogu2020protamine}. In addition, we have set one binding site every 10 bp, which is the pitch of the DNA, making bending in one direction more likely. Future simulations could relax this requirement.

The second consideration is to determine the bend angle. Previous theory of DNA bending by small cations shows bending of $20^{\circ}-40^{\circ}$. However, molecular dynamics simulations of protamine-DNA complexes show different DNA bend angles \cite{mukherjee2021protamine}. Thus, we use an implementation where the condensing agent takes advantage of thermal fluctuations in the DNA to aid bending. Specifically, if the initial thermal bend $\beta_i$ is less than a threshold angle $\beta_\text{thresh}$, then we set the final bend angle for the region as $\beta_{\text{final}, \, i} = \beta_\text{thresh}$. On the other hand, if $\beta_i$ is greater than $\beta_\text{thresh}$, then we take advantage of the greater bend angle at this site and set $\beta_{\text{final}, \, i} = \beta_i$. For the case of protamine-DNA, we set $\beta_\text{thresh} = 47^{\circ}$. This angle is more than twice the standard deviation $\sigma\approx21^\circ$ for thermal fluctuations (Eq.~\ref{eq:BetaPDF}). Thus, the final bend angle $\beta_{\text{final}, \, i}$ is approximately constant at 47$^{\circ}$.

The third consideration is to determine if there is a maximum bend angle. If we assume that the condensing agent is a sphere, then there will be a maximum bend angle due to steric considerations \cite{rouzina1998dna}. To find this bend angle, we use Eq.~\ref{eq:r} which determines $r$, the separation between the center of the condensing agent and the center of the phosphate on the DNA backbone. This separation has a minimum value
\begin{equation}
r_\text{min} = r_\text{agent} + r_\text{phos}, 
\end{equation}

\noindent where $r_\text{phos}$ is the effective radius of the phosphate (0.29 nm) and $r_\text{agent}$ is the effective radius of the condensing agent. Thus, we can rearrange Eq.~\ref{eq:r} to find the maximum sterically allowed bend angle
\begin{equation} \label{eq:BetaMax}
\beta_\text{max} = \frac{2}{R_\text{DNA}} \left(r_0 - r_\text{min} \right).
\end{equation}

\noindent For cobalt hexaammine (III), $r_\text{agent}$ is 0.3 nm, which yields an $r_\text{min}$ of 0.59 nm and a maximum bend angle $\beta_\text{max}$ of $47^\circ$. Likewise, spermine and spermidine have maximum bend angles of $45^\circ$ \cite{rouzina1998dna}. Protamine is a disordered protein and does not have a spherical structure \cite{shadman2022exploring}. Thus, we might not expect a maximum bend angle for protamine. Indeed, molecular dynamics simulations of protamine-DNA complexes show DNA bending of up to $\sim120^\circ$ \cite{mukherjee2021protamine}. We therefore did not encode a maximum bend angle.

\subsection*{Introducing condensing agents: probability of binding}

Finally, we calculate the probability that a condensing agent binds at the $i^\text{th}$ site, given the energies for binding and bending. To do this, we first calculate the total energy at the $i^\text{th}$ binding site,
\begin{gather}
U_{\text{tot}, \, i} = U_{\text{bind}, \, i}+ U_{\text{bend}, \, i},
\end{gather}

\noindent using Eq.~\ref{eq:UBindSimplified} and Eq.~\ref{eq:UBendSimplified}. Then, we use $U_{\text{tot}, \, i}$ as the energy term in the Boltzmann distribution to calculate the binding probability at the $i^\text{th}$ site,
\begin{equation} \label{eq:BindingProb}
p_i = \frac{1}{Z} e^{-\frac{U_{\text{tot}, \, i}}{k_B T}}.
\end{equation}

\noindent This probability is used as a weight in determining binding location (Fig. \ref{fig:Methods}C).

\subsection*{Checking for loop formation}

Once we have a DNA contour with thermal fluctuations and bound condensing agent, we need to check for loop formation. Loop formation could happen in two ways (Fig. \ref{fig:Methods}D). First, a loop could form if there is a DNA ``crossover'', which is a location where the DNA overlaps itself. Second, a loop could form if two locations on the DNA are ``close'' enough to each other, perhaps using thermal fluctuations to close the rest of the distance. We call this ``thermal loop closing''. 

Thermal loop closing works as follows. We create a loop in a DNA molecule if (1) the separation $r_{ij}$ between two locations $i$ and $j$ on the DNA is less then some threshold distance $k_\text{planar} L_P$ for some factor $k_\text{planar}$, and (2) the separation $|s_i - s_j|$ between the two locations along the DNA contour is greater than some threshold distance $k_\text{linear} L_P$ for some factor $k_\text{linear}$. The first criterion selects only ``close'' segments to be stabilized by protamine, while the second criterion is meant to exclude regions on the DNA that are adjacent to one another.

\section*{Materials and Methods}

\subsection*{Simulation}

We write MATLAB code to implement the bind-and-bend model. First, we produce simulated DNA molecules with spontaneous thermal fluctuations using steps \ref{item:thermal_first}-\ref{item:thermal_last} below.
\begin{enumerate}
 \item Discretize the DNA into $n$ regions with one binding site each.\label{item:thermal_first}
 \item For each of the $n$ binding sites, determine the spontaneous thermal bend $\beta_i$ at that binding site by sampling angles from the probability distribution in Eq.~\ref{eq:BetaPDF} with the MATLAB function ``normrnd''.
 \item Bend the DNA at each binding site to obtain a piece-wise linear representation of the DNA contour.
 \item Apply a cubic smoothing spline using the MATLAB function ``csaps'' to smooth the DNA contour. This is done to remove abrupt, unrealistic changes in the direction of the DNA contour. \label{item:thermal_last}

\end{enumerate}

\noindent Next, we add condensing agents to the DNA and update the DNA contours using steps \ref{item:bend_first}-\ref{item:bend_last}. 
\begin{enumerate}
\setcounter{enumi}{4}
\item For each binding site, sum Eqs.~\ref{eq:UBindSimplified} and \ref{eq:UBendSimplified} to compute the total energy change upon condensing agent binding.\label{item:bend_first}
\item For each binding site, apply the Boltzmann distribution to the total energy from step \ref{item:bend_first} to compute the relative probability in Eq.~\ref{eq:BindingProb} that the condensing agent binds at that site.
\item Use these relative probabilities as weights in the MATLAB function ``randsample'' to determine where the next condensing agent binds.
\item Bend the region of DNA where the condensing agent has bound, and update the charge distribution of the DNA. Recompute the DNA contour. \label{item:bend_secondlast}

\item Repeat steps \ref{item:bend_first} through \ref{item:bend_secondlast} for each condensing agent that binds to the DNA. The number of iterations is the number of condensing agents, $N_\text{agents}$. \label{item:bend_last}
\end{enumerate}

\noindent After steps \ref{item:thermal_first}-\ref{item:bend_last}, we have a single DNA contour that represents one simulated DNA molecule. 

Parameters and constants for the simulation are listed in Table S1 and Table S2 respectively. The DNA polymer is defined by the contour length $L_C$, the persistence length $L_P$, and the linear charge density $\lambda$. The condensing agent is defined by the size of the region $L$, the number of condensing agents $N_\text{agents}$, the charge of the condensing agent $q_\text{agent}$, and the smallest bend angle $\beta_\text{thresh}$. There are three terms to account for the effective electrostatic screening of charges, namely, $p_\text{agent}$, $p_\text{DNA}$, and $p_\text{site}$. There are two factors that determine if DNA regions $i$ and $j$ are close enough for thermal loop closing, namely, $k_\text{planar}$ (for the condition $r_{ij} < k_\text{planar} L_P$) and $k_\text{linear}$ (for the condition $|s_j-s_i| > k_\text{linear} L_P$). 

To generate a simulated DNA molecule with no condensing agents, we follow steps \ref{item:thermal_first}-\ref{item:thermal_last}. We repeat the process multiple times to produce a data set of simulated DNA molecules without condensing agents.

To generate a simulated DNA molecule with condensing agents, we follow steps \ref{item:thermal_first}-\ref{item:bend_last}. We repeat the process multiple times to produce a data set of simulated DNA molecules with condensing agents. Every 10th molecule, we add only half the number of condensing agents in order to simulate molecules with a distribution of bound condensing agents.

\subsection*{Classifying simulated molecules}

We classify molecules as single loops or flowers using an additional MATLAB code. In this code, we first check for segment intersection (see Supporting Material, ``Determining Segment Intersection Points'', and Fig. S1) using a known algorithm \cite{cormen2009introduction}. If we find segment intersection, we identify the loop start site at the crossover location. Otherwise, we check for distal regions of the DNA that are ``close'' to each other to identify loop formation by thermal loop closing. Distal regions are considered ``close'' if $r_{ij} < k_\text{planar} L_P$ and $|s_j-s_i| > k_\text{linear} L_P$. We select the location with the smallest planar separation $r_{ij}$ and identify a loop start site at that location.

To classify a molecule as a 2-loop flower or 3-loop flower, we use the following rules:
\begin{itemize}
 \item The molecule must have two or three distinct loops.
 \item All of the loops must be bent in phase. That is, traveling along the DNA contour in the same direction, all of the loops must have the same orientation (i.e., clockwise or counterclockwise).
 \item Loops must be ``close enough'' to each other. Specifically, the start sites of the loops must lie within a distance $\frac{L_P}{2}$ of each other as measured along the DNA contour. Or, for a loop that is closed by thermal loop closing at a DNA end, the end must lie within a planar distance $\frac{L_P}{2}$ of a loop start site. 
\end{itemize}
\noindent Some examples of 2-loop flowers, 3-loop flowers, or molecules with multiple loops are shown in Fig. S2 and Fig. S3.

\subsection*{Analyzing simulated molecules}
To calculate the radius of curvature, molecules are first smoothed with a cubic smoothing spline using the ``csaps'' function in MATLAB. Then, we use the ``LineCurvature2D'' function \cite{kroon2011curvature} to calculate the radius of curvature for each region.

For molecules with a single loop, we measure the circumference $c$ and the distance to the start site $s_s$. The circumference is the distance $|s_j-s_i|$ along the contour of the loop, where the $i^\text{th}$ and $j^\text{th}$ region are at the loop crossover point or at the smallest planar separation $r_{ij}$ for thermal loop closing. The start site distance is the smaller of $s_i$ or $L_C-s_j$, where the $i^\text{th}$ region occurs before the $j^\text{th}$ region.

For flowers, we measure by hand the flower start site as the distance along the contour from the nearest DNA end to the flower center. The flower center is defined as the midpoint between the most extreme loop start sites. For flowers that have one loop with thermal loop closing, we set $s_s = 0$.

\subsection*{Preparing DNA constructs and protamine}
We use template DNA from bacteriophage lambda (N3011; New England Biolabs, Ipswich, MA) to produce DNA constructs of lengths 309 bp, 639 bp, 1170 bp, and 3003 bp. Polymerase chain reaction using customized oligonucleotide primers (Integrated DNA Technologies, Coralville, IA) and an LA Taq DNA polymerase (RR004; TaKaRa Bio, Kusatsu, Japan) generated the various DNA lengths. Gel electrophoresis determined if the DNA had been amplified properly, and DNA extraction was performed using an extraction kit (QIAquick PCR Purification Kit; Qiagen, Hilden, Germany). A nanodrop spectrophotometer (NanoDrop Lite; Thermo Fisher Scientific, Waltham, MA) assessed the concentration and purity of the DNA samples, and samples with A260/A280 purity ratios below 1.7 were not used. 

Protamine from salmon (P4005; Sigma-Aldrich, Saint Louis, MO) was diluted in water to 30 {\textmu}M and stored in 30 {\textmu}L aliquots at -20℃. To check for aggregation, we used an AFM to image a solution of 10 {\textmu}M protamine and 1 mM magnesium acetate bound to a mica slide \cite{mcmillan_dna_nodate}. We did not see any protamine aggregates. 

\subsection*{Preparing AFM slides}
Procedures were the same as in previous experiments \cite{ukogu2020protamine, mcmillan2021dna, mcmillan_dna_nodate}. Briefly, we prepared AFM samples by attaching 10-mm-diameter ruby muscovite mica slides (grade V1; Ted Pella, Redding, CA) to a metal disc. Tape was used to clean the surface of the mica. Then, a DNA solution consisting of 0.2 ng/{\textmu}L DNA, 1-2 mM magnesium acetate, and protamine (concentrations of 0.0-5.0 {\textmu}M) was prepared. The solution was pipetted onto the slide, rinsed with 1 mL of deionized water, and dried with nitrogen. This procedure was repeated in order to obtain 1-5 depositions of DNA solution on each mica slide. 

For the control, the DNA solution consisted of 1-2 mM magnesium acetate and 1.0 ng/{\textmu}L DNA. The solution was deposited onto the mica slide. After waiting for 30 seconds, we rinsed with 1 mL of deionized water and dried with nitrogen. 

\subsection*{Imaging AFM slides}

AFM samples were imaged with a Dimension 3000 AFM (Digital Instruments, Tonawanda, NY) or an MFP-3D AFM (Asylum). AFM tips (PPP-XYNCSTR-model; Nanosensors, Neuchatel, Switzerland; Parameters: resonant frequency = 150 kHz, force constant = 7.4 N/m, length = 150 {\textmu}m, tip radius < 7 nm) were set to tapping mode. The scan rate was 1-4 Hz. AFM resolution in $z$ is $\sim$0.2 nm \cite{devenica2016biophysical} and the lateral resolution is limited by the tip radius. Image size was 1-5 {\textmu}m square with either 256 or 512 pixels per line. 

\subsection*{Analyzing AFM images}

We processed images with Gwyddion. Rows were aligned with a 5th-degree polynomial. High-frequency oscillations were removed with a fast Fourier transform filter. Scars were removed. We cropped images of single DNA molecules for further use. The molecules were flat on the surface ($<$0.5 nm) and were at least 1 pixel apart from other molecules. 

These singlets, along with singlets from previous experiments \cite{ukogu2020protamine, mcmillan2021dna, mcmillan_dna_nodate} were analyzed. For each DNA singlet, we determined the number of loops. Molecules with multiple loops were labeled as flowers. We took two perpendicular profiles of each loop and measured the diameter of the loop in each profile before averaging to obtain $d$. The circumference was calculated as $c=\pi d$. We measured the distance to the start site $s_s$ and contour length $L_C$ of each DNA molecule, using previous methods \cite{mcmillan2021dna}. The distance to the start site is defined as the arc length of the shortest tail. We then calculated a fractional value for the start site distance $s_s/L_C$. Instead of plotting a histogram of these values, we plot a probability density using kernel density estimation with the MATLAB function ``ksdensity''.

To calculate the radius of curvature, we use a customized program to extract the 309-bp DNA contours from the image using the ``regionprops'' command in Matlab, as described previously \cite{ukogu2020protamine}. Then, we calculate the radius of curvature at each point in the contour for each DNA singlet using the ``LineCurvature2D'' function in Matlab \cite{kroon2011curvature}. 

\section*{Results}

\subsection*{Model captures spontaneous loop formation}

We first tested the portion of the bind-and-bend model used to generate spontaneous thermal fluctuations (steps \ref{item:thermal_first}-\ref{item:thermal_last}). If this portion of the model is working properly, simulated DNA molecules much longer than a few persistence lengths ($L_C \gg L_P$) should form spontaneous loops.

\begin{figure}[!htb]
\centering
\includegraphics[width=0.92\linewidth]{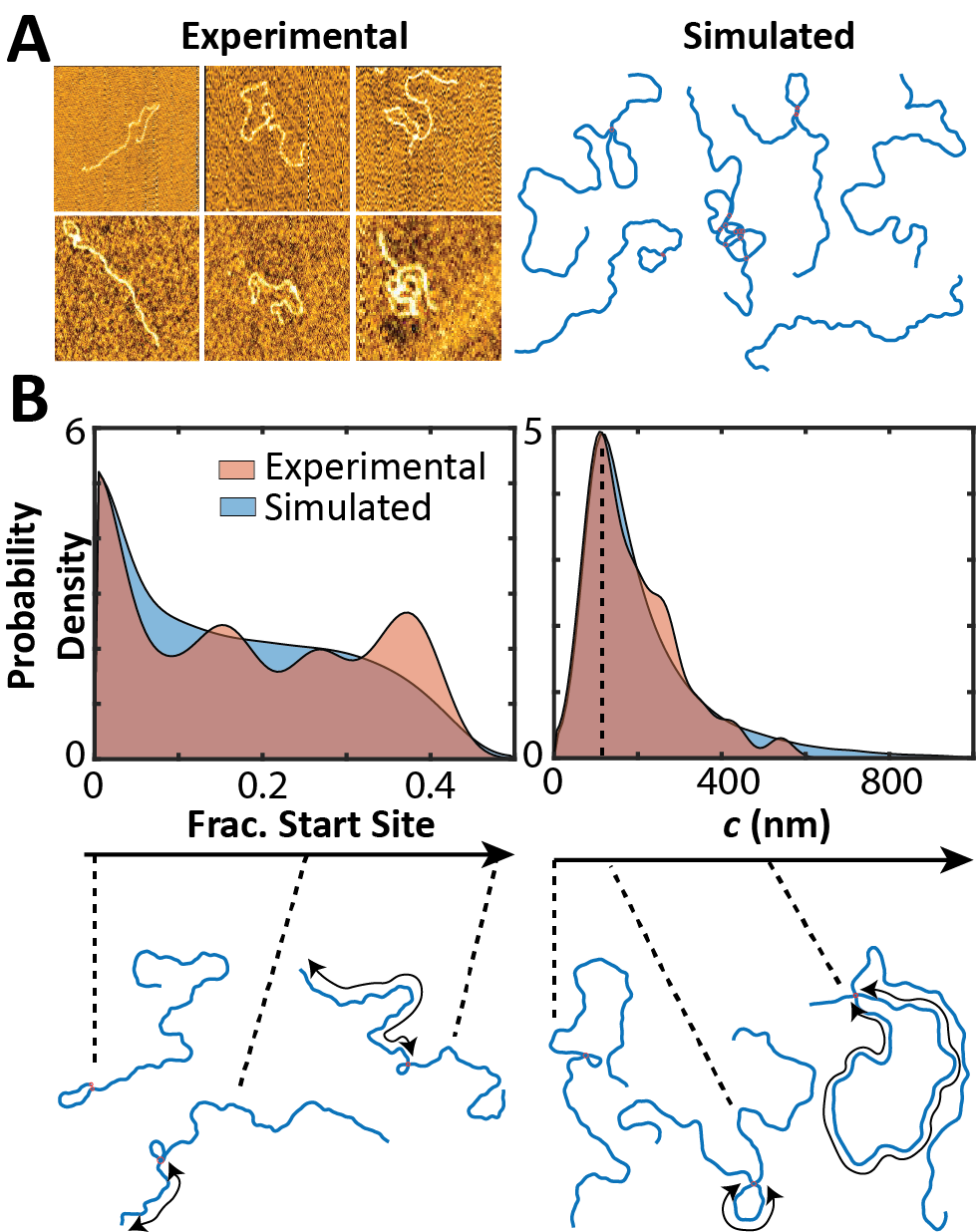}
\caption{\textbf{Model accurately predicts spontaneous loop formation.} A) Experimental AFM images and simulated contours of 3003-bp-long DNA molecules. B) We isolated the single-looped DNA molecules from the experimental (\emph{red}, $N = 44$) and simulated (\emph{blue}, $N = 23920$) data sets. For each molecule, we computed the fractional distance to the loop start site $s_s/L_C$ as the length of the shorter tail and the loop circumference $c$. Simulated molecules with varying start site and loop circumference shown for illustration.}
\label{fig:Spontaneous}
\end{figure}

To test this portion of the model (Fig. \ref{fig:Spontaneous}), we simulated 100,000 DNA molecules of contour length $L_C= 3003$ bp with the parameters in Row 1 of Table S1. We then analyzed AFM images of 80 DNA molecules of the same length immobilized on the surface with magnesium acetate. These molecules equilibrate on the surface and are in a random conformation due to thermal fluctuations \cite{rivetti1996scanning}. Qualitatively, the simulated molecules look similar to the actual molecules in the amount of bending, overall shape, and conformational variety. To quantify the data, we analyzed molecules with a single loop (number of molecules, $N = 23920$ in the simulation and $N = 44$ in the experimental data) and computed the distance to the start site $s_s$, and loop circumference $c$. We fractionalize the distance to the start site by dividing by the contour length. We plot the probability densities for $s_s/L_C$ and $c$, and we compare the simulated and experimental data sets.

In the fractional start site distance distribution, the model captures all of the features of the experimental data, including a drop off point at $s_s/L_C \approx 0.45$, a relatively flat distribution for $0.1 \lesssim s_s/L_C \lesssim 0.3$, and a peak at $s_s = 0$. The drop off point at $s_s/L_C \approx 0.45$ is due to the geometry of the molecule. Indeed, the maximum distance to the start site would occur for a loop ($c= 300$ bp) in the very middle of the DNA (1500 bp), producing a tail of 1350 bp ($s_s/L_C \approx 0.45$). The roughly uniform probability density for $0.1 \lesssim s_s/L_C \lesssim 0.3$ indicates that loops are randomly forming along the DNA without any bias. The small variations we see in the experimental data are due to statistical errors. Finally, the peak at $s_s = 0$ has been observed previously and is due to a statistical end-effect \cite{mcmillan2021dna}. 

Specifically, the statistical end-effect is because fluctuations near the DNA end always cause loops to form at the DNA end rather than in the middle of the DNA. So even though the probability for a thermal fluctuation is evenly distributed along the DNA, the probability for a loop to form at the DNA end is higher than the probability for a loop to form in the middle of the DNA. However, to observe loops occurring at the DNA ends in the simulation, we need to add in thermal loop closing. Thermal loop closing ensures that loops at the DNA end that are ''almost closed'' (with $r_{ij} < k_\text{planar}L_P$) are counted (since the AFM tip has a radius of 7 nm, loops that are almost closed in the experimental data are counted as loops). If we apply the model with $k_\text{planar}$ equal to zero (parameters in Row 2 of Table S1), this removes thermal loop closing and the peak at $s_s = 0$ disappears (Fig. S4). Thus, we only detect enough loops at $s_s = 0$ in our simulation if we add in thermal loop closing.

In the loop circumference distribution, the model captures the height and location of the peak at $c \approx 120$ nm (350 bp). It also captures the right skew of the distribution, with a circumference range of 50-500 nm (150-1500 bp). This range is much larger than the range of protamine-induced loops, which is 200-300 bp \cite{ukogu2020protamine}.

\subsection*{Model reproduces protamine-induced DNA bending}

Next, we tested the portion of the bind-and-bend model used to generate condensing agent binding and DNA bending (steps \ref{item:bend_first}-\ref{item:bend_last}). If this portion of the model is correct, then the simulation should reproduce the shape of short ($L_C \sim 2 L_P$) DNA molecules that are exposed to protamine. At this DNA length, molecules will exhibit DNA bending and have very little conformational variety due to thermal fluctuations.

\begin{figure}[!htb]
\centering
\includegraphics[width=0.92\linewidth]{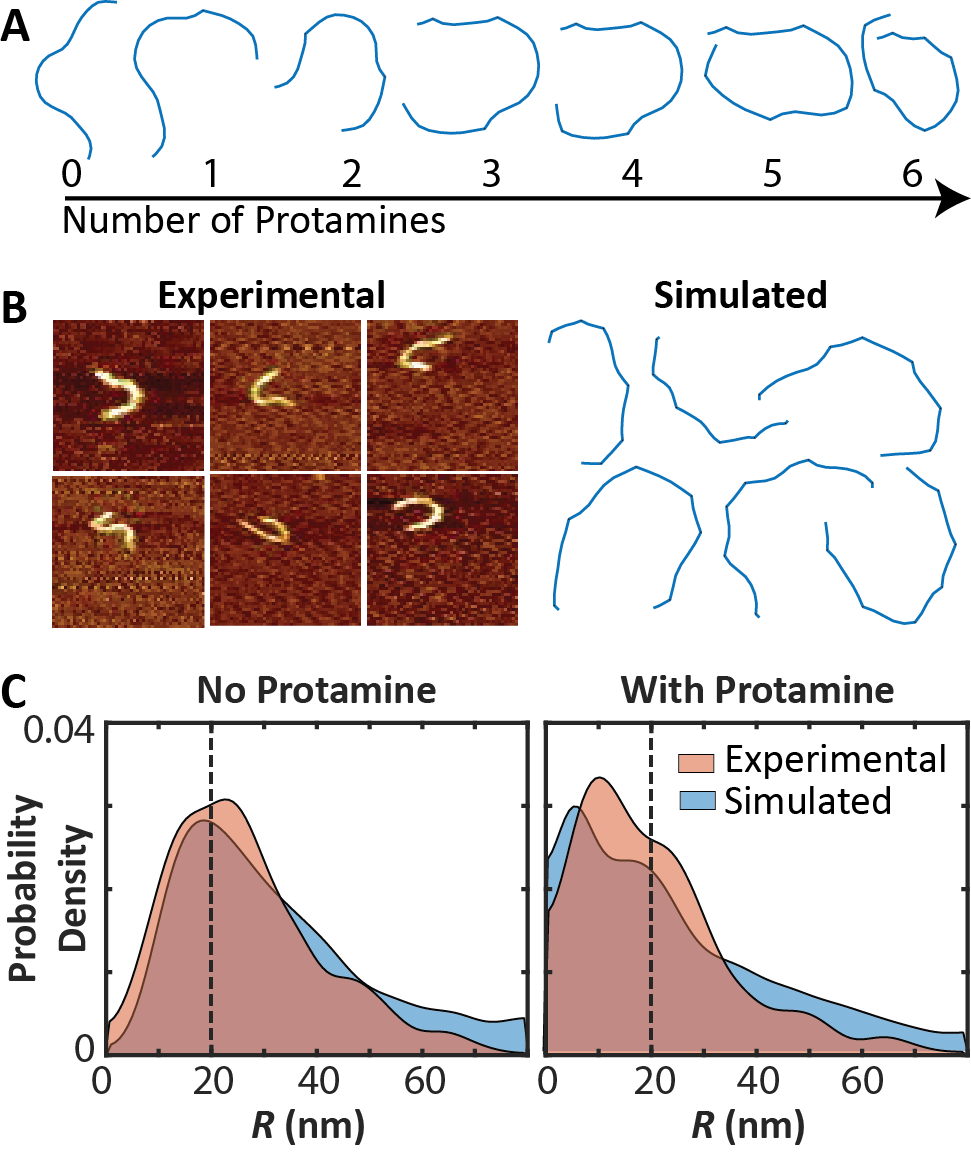}
\caption{\textbf{Model simulates protamine-induced DNA bending.} A) Step-wise addition of protamine to the DNA induces bending. B) Experimental AFM images and simulated DNA contours of 309-bp-long DNA molecules. C) Distributions of the radii of curvature of molecules with and without protamine. Molecules with protamine have a smaller radius of curvature ($R < 20$ nm) than molecules without protamine.}
\label{fig:105Bending}
\end{figure}

To test this portion of the model (Fig. \ref{fig:105Bending}), we simulated 100 DNA molecules of contour length $L_C = 309$ bp in the absence ($N_\text{agents}= 0$) and presence ($N_\text{agents}= 6$) of protamine. Parameters used in the simulation are in Rows 3 and 4 of Table S1, respectively. We compared these simulated molecules to actual images of immobilized, 309-bp-length DNA molecules in 0 {\textmu}M ($N$ = 100) and 0.2 {\textmu}M protamine ($N$ = 94). When we look at the data, we see that both the simulated and experimental molecules in the presence of protamine have the characteristic ``C'' shape of DNA bending and that the bend angle is similar. One difference is that bending in the simulation is all in-phase; bends always happen to one side (as expected). In the experimental data, 20$\%$ of molecules have bending in both directions. Future simulations could update this feature.

To quantify the data, we measured the radius of curvature $R$ at $\sim$30 locations along the DNA contour (Fig. \ref{fig:105Bending}C). In the absence of protamine, both the simulation and the experimental data produce a peak at $\sim$20 nm, indicating that the model is capturing the thermal fluctuations of the DNA. Thermal fluctuations in the DNA should give curvatures close to about half the persistence length ($L_P/2= 75$ bp or 25 nm). In the presence of protamine, both the experimental and simulated data still have a peak at 20 nm, but now there is another peak at a lower $R$ value due to DNA bending by protamine. This peak is at about 10 nm in both data sets, which corresponds to an average bend angle of $\sim$20$^\circ$ (specifically, this angle in radians is found by dividing the arc length of a region by the radius of curvature). However, the bend angle by protamine in the simulation is actually $\beta_\text{thresh} = 47^\circ$, which should correspond to a 4 nm radius of curvature. The reason why we observe $\sim$10 nm and not 4 nm is that the csaps program we use to create the DNA contour smooths the data, averaging curvature due to protamine bending (4 nm) and thermal fluctuations (20 nm).

To quantify the data further, we also measure the decay of the tangent-tangent correlation and how the mean squared displacement along the contour varies with contour length (Fig. S5) using the Easyworm software \cite{lamour2014easyworm}. See Supporting Material ``DNA Bending by protamine'' for more information. In both measurements, the simulated and experimental molecules show similar observables (radius of curvature of 10-20 nm).

Thus, we find that our simulation reproduces experimental protamine bending. However, we previously estimated bending by protamine to be about $20^\circ$ \cite{ukogu2020protamine}. Here, we find that in order for our simulated data to match our experimental data, we need a much larger bend angle of $\beta_\text{thresh} = 47^\circ$. The reason we need a much larger bend angle is that some sites in the DNA remain unoccupied. In our previous calculation \cite{ukogu2020protamine}, we assumed that every binding site was occupied. If some binding sites remain unoccupied or if some protamine molecules do not bend the DNA, then the bend angle for protamine molecules that do bend the DNA would need to be much higher. Since molecular dynamics simulations show protamine-DNA complexes without any DNA bending and other complexes with large DNA bending of up to $120^\circ$ \cite{mukherjee2021protamine}, it is likely that we underestimated the bend angle for those protamines that bind and bend the DNA. Indeed, a higher bend angle for protamine might be advantageous given protamine's role to condense the DNA within sperm cells in minutes \cite{balhorn_protamine_2007, vilfan_formation_2004}.

\subsection*{Model fits protamine-induced loop formation}

\begin{figure}[!htb]
\centering
\includegraphics[width=0.92\linewidth]{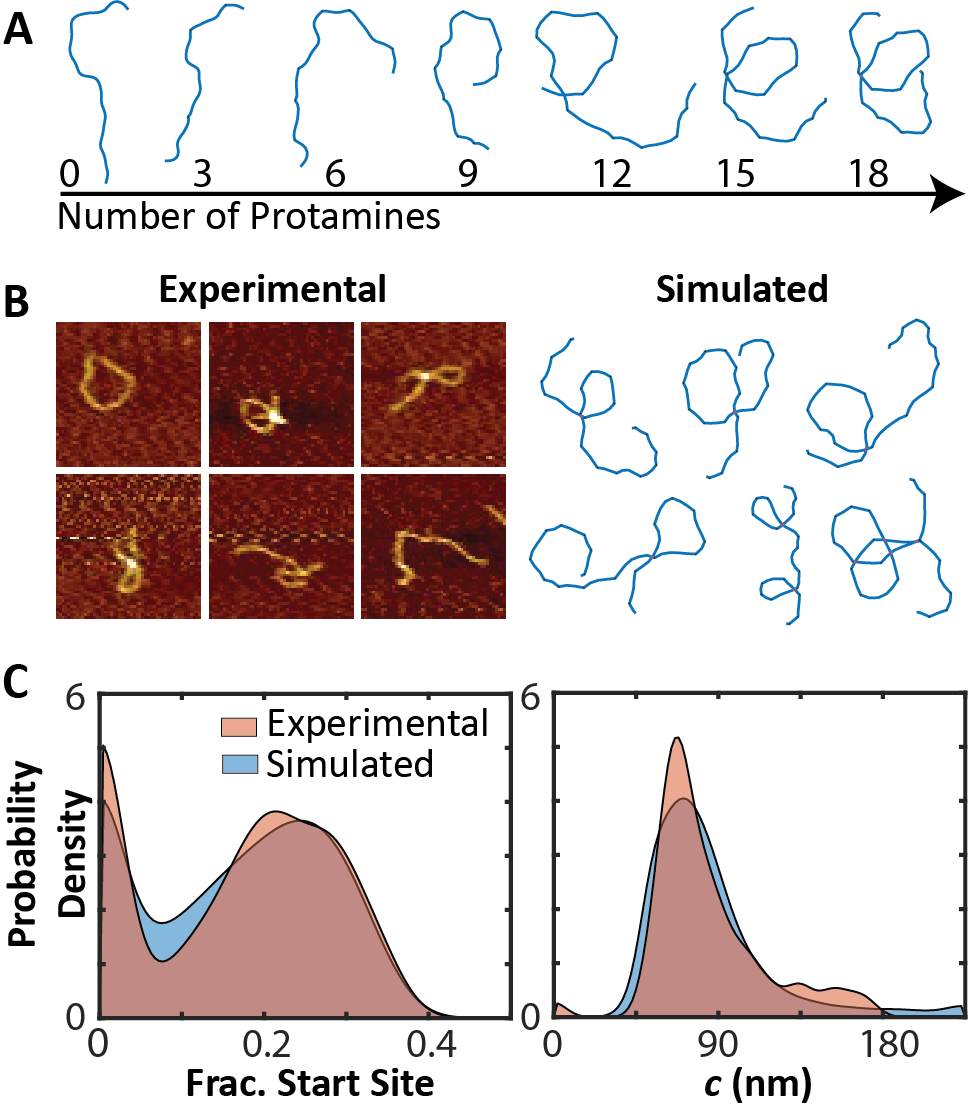}
\caption{\textbf{Model accurately simulates protamine-induced loop formation.} A) An unlooped 639-bp DNA molecule is folded by the addition of protamine molecules to form a single loop. B) Experimental data from AFM images of 639-bp DNA in the presence of 0.2-5.0 {\textmu}M protamine ($N= 77$) and simulated contours ($N \sim 22000$) for the same DNA length with $N_\text{agents}= 18$ molecules. C) The experimental (red) and simulated (blue) distributions of fractional start site distance and loop circumference.}
\label{fig:SingleLoops}
\end{figure}

Finally, we tested the full bind-and-bend model (steps \ref{item:thermal_first}-\ref{item:bend_last}) to see if the model reproduces looping of DNA by protamine (Fig. \ref{fig:SingleLoops}). To test the model, we simulated 100,000 molecules which were 639 bp long (parameters in Row 5 of Table S1). We compared this simulated data to experimental AFM images of 639 bp DNA molecules that were immobilized on the surface by magnesium acetate in the presence of 0.2-5.0 {\textmu}M protamine. We then repeated these measurements for 1170 bp DNA (Fig. S6). Qualitatively, we see that the experimental and simulated data agree. Both the simulated and experimental molecules fold into structures with 0-2 loops and have a similar loop size and shape. 

To quantify the data, we isolated the molecules with single loops ($N= 77$ for the experimental data and $N \sim 22000$ for the simulated data) and computed the circumference $c$ and fractional distance to the start site $s_s/L_C$ for each molecule. We then plotted the probability densities for the experimental and simulated data. Both distributions of loop circumference have a peak at $\sim70$ nm ($\sim200$ bp), indicating a radius of curvature of about 10 nm. This matches the protamine-induced bending we observed in Fig. \ref{fig:105Bending} and our previous measurements \cite{ukogu2020protamine}. In addition, both distributions are right-skewed, meaning that there are many loops that have a circumference larger than 70 nm. In the simulation, these loops are partially or fully formed by spontaneous thermal fluctuations. Loops formed by spontaneous thermal fluctuations have a peak circumference at 120 nm and can be as large as 500 nm (Fig. \ref{fig:Spontaneous}B).

In addition, both the experimental and simulated data produced similar distributions of the fractional start site distance $s_s/L_C$. In particular, both distributions had a large peak at $s_s/L_C = 0$ and $s_s/L_C \approx 0.25$, suggesting that the simulation is capturing the DNA looping process. The peak at $s_s/L_C = 0$ indicates that there is a significant proportion of loops that form at the end of the DNA due to thermal loop closing. To verify this, we ran the same simulations without thermal loop-closing (i.e., with $k_\text{planar} = 0$, Row 6 of Table S1) and found that the peak at $s_s/L_C = 0$ disappears (Fig. S4). 

Given our agreement between simulation and experimental data, we can now look to the simulation to see how protamine is using bind-and-bend to fold the DNA into a loop. Before protamine is bound, most DNA molecules are unlooped and fairly straight (Fig. \ref{fig:SingleLoops}A and Fig. S7). Protamine molecules then bind to the DNA, with binding occurring first at locations where there is a high concentration of DNA (regions with DNA crossovers or high curvature), before binding at locations that are spread out along the DNA. This creates bending all over the molecule. Eventually, this bending leads to the formation of a loop. To see this clearly, we set the simulation to stop as soon as a loop is formed. Contours at this point in the simulation (Fig. S8) have loops with large circumferences ($c \sim L_C$) and small fractional start sites ($s_s/L_C \sim 0$). As more protamines bind the DNA, the protamines bind and bend the DNA within the loop, decreasing the loop circumference and increasing the distance to the start site. This drives the peak in the loop circumference distribution away from $c = L_C$ towards $c = 0$ and the second peak in the factional start site distribution away from $s_s/L_C = 0$ towards $s_s/L_C = 0.5$. However, at some point, it becomes unfavorable for additional protamines to bind within the loop, making further decreases in $c$ and increases in $s_s/L_C$ unlikely. Thus, the peak in the loop circumference and the second peak in the fractional start site distribution approach a preferred value set by the DNA bend angle and electrostatics. Here that preferred value is a loop circumference of 70 nm and for a second peak in the fractional start site distribution of $s_s/L_C \sim 0.25$. 

Other condensing agent-DNA systems may have smaller or larger loop circumferences. Smaller loop circumferences could occur with a higher DNA bend angle or if the effective charge on the DNA is such that there is more binding within the loop.

We note that the bind-and-bend model does not account for DNA-DNA interactions. Here the size of the loop is not set by protamine stabilizing the DNA crossover location. Instead the size of the loop is completely set by DNA bending and the electrostatic properties of the system. Indeed, our model predicts that if protamine DNA-DNA interactions are present, they are not stable, since then the fractional start site distribution would be closer to $s_s/L_C = 0$ and the loop circumference would be larger. Why does our model not have to account for DNA-DNA interactions? One reason could be that folding in solution occurs in 3D, not in 2D like our model. In 3D, the DNA would be much less likely to have a DNA crossover point. Another reason could be that molecules with less than 2 DNA-DNA interactions are known to be unstable \cite{van2010visualizing}. Perhaps this instability allows us to neglect DNA-DNA interactions when looking at loop formation.

\subsection*{Model predicts flowers}

\begin{figure}[!htb]
\centering
\includegraphics[width=0.92\linewidth]{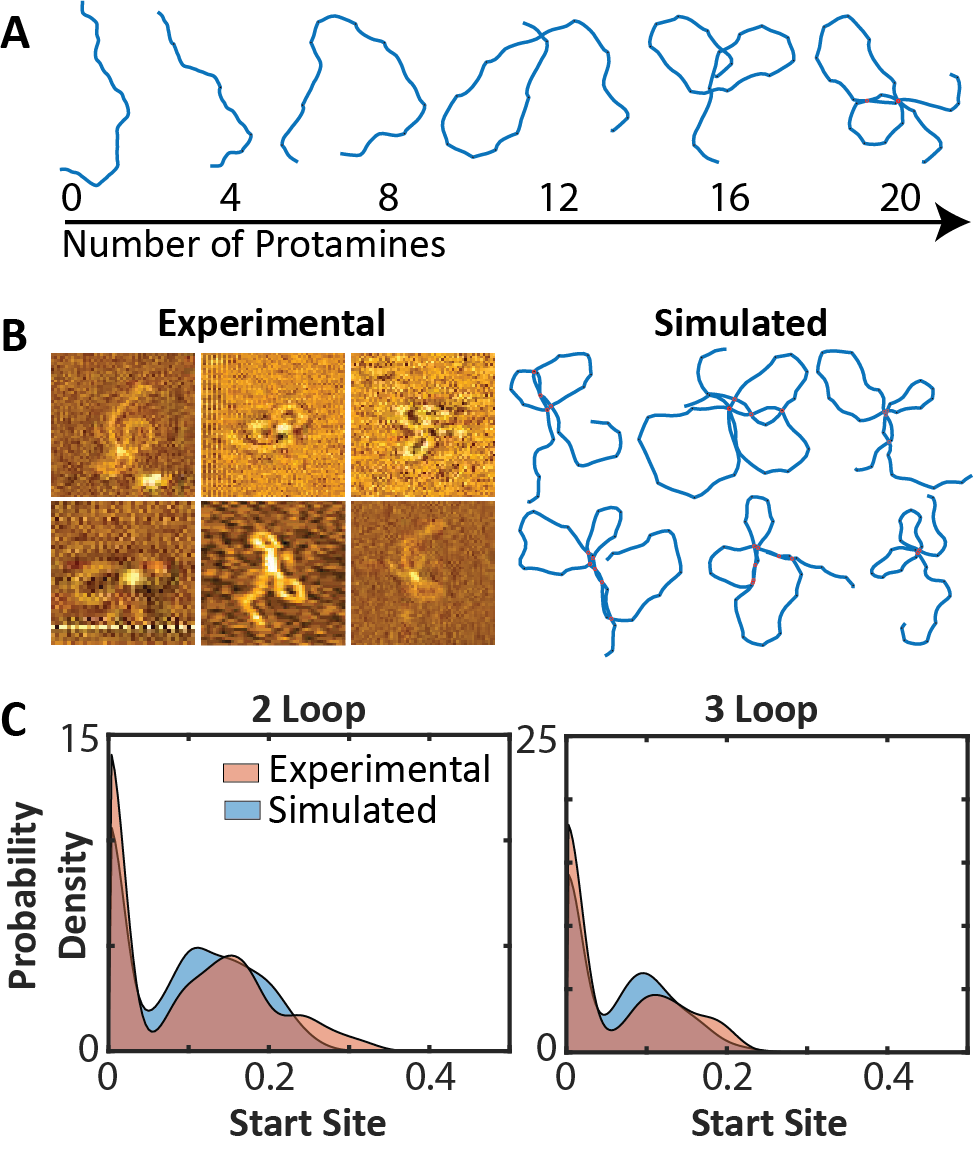}
\caption{\textbf{Model accurately predicts protamine-induced flower formation.} A) An unlooped 1170-bp DNA molecule is folded progressively by protamine molecules to form a 2-looped flower. B) We identified AFM images of DNA flowers (left) and isolated flower-shaped molecules from our simulation (right). C) We further classified flowers as double-looped ($N= 78$ experimental molecules and $N= 478$ simulated molecules) or triple-looped ($N= 42$ experimental molecules and $N= 180$ simulated molecules) and computed the distribution of the fractional distance to the start site for both the experimental (red) and simulated (blue) data sets.}
\label{fig:Flowers}
\end{figure}

Next, we applied our model to investigate the formation of flower-shaped DNA molecules (Fig. \ref{fig:Flowers}). A DNA flower is a multilooped molecule where the loops share a common crossover point. Would the bind-and-bend model predict such structures given protamine-induced bending?

We used our simulation to produce $2,000$ DNA contours of length 1170 bp with high amounts of condensing agent (with $N_\text{agents}$ equal to 20 or 25). Parameters are in Row 9 and Row 10 of Table S1, respectively. We observe that the simulation does produce flowers! However, we observe that only a third of multilooped structures are flowers in the simulation, compared to experimental reports of 87$\%$ \cite{mcmillan_dna_nodate}.

To make a quantitative comparison (Fig. \ref{fig:Flowers}), we identify all of the flowers in the simulation ($N= 478$ with two loops and $N= 180$ with three loops) and compare these simulated flowers to actual AFM images of flowers ($N= 78$ with two loops and $N= 42$ with three loops) made from 1170 bp DNA at 0.2-2 {\textmu}M protamine. We measure the distance to the start site for every flower, that is, the shortest distance from the DNA end to the flower's common crossover point. When we plot the distribution of start sites for the experimental and simulated data, we see that the distributions overlap. The model predicts the location and height of the first peak in the fractional start site distance of $s_s/L_C = 0$ for both 2-loop and 3-loop flowers. This peak is due to flowers that have at least one loop with thermal loop closing. In addition, the model captures the location and height of the second peak at $s_s/L_C \approx 0.15$ and $s_s/L_C \approx 0.1$ in the 2-loop and 3-loop fractional start site distributions, respectively. This peak in the fractional start site distance is due to a balance between protamine binding within loops in the flower when the loop circumferences are large (>70 nm) and protamine binding outside of the loops when the loop circumferences are small (<70 nm). The model also correctly predicts that the distance to the peak in the 2-loop fractional start site distribution is greater than the distance to the peak in the 3-loop fractional start site distribution. Finally, the model predicts that the probability density falls off at $s_s/L_C$ of 0.3 or 0.2 for 2-loop or 3-loop flowers, respectively, due to the geometric constraints of the molecule.

\section*{Discussion}
\subsection*{Model summary}
One method for creating a DNA loop is bind-and-bend. Here, we theoretically and computationally model bind-and-bend using the following steps: i) discretize the DNA into regions, ii) select a bending angle $\beta_i$ for each region due to spontaneous thermal fluctuations \cite{rouzina1998dna}, iii) select a binding site for each condensing agent based on a Boltzmann distribution of the bending and binding energies in that region \cite{rouzina1998dna}, iv) bend the DNA at the binding site, and v) repeat the procedure for each condensing agent that binds to the DNA. This simulation reproduces DNA looping due to spontaneous thermal fluctuations, DNA bending due to protamine, DNA looping due to protamine, and the folding of the DNA into multilooped flowers. In particular, the simulation captures the changes in DNA curvature due to both protamine-induced bending and spontaneous thermal fluctuations, the large number of loops that form at the DNA ends, the bias in the loop start site location that occurs at a fractional start site distance $s_s/L_C$ of about 0.25 (for single loops of 639-bp DNA), and the small circumference of the DNA loops of 70 nm (200 bp). 

Using this model, we are able to describe how protamine uses bind-and-bend to fold the DNA into a loop. We observe that protamine molecules bind all along the DNA, bending the DNA at each location. Eventually, through a combination of spontaneous thermal fluctuations and protamine-induced bending, the DNA curves around into a loop. The bending then continues to increase the DNA curvature until electrostatics limits further protamine molecules from binding within the loop. This sets the radius of curvature for protamine-DNA at $\sim$10 nm.

\subsection*{Model advantages and limitations}
There are some advantages of this model. First, the model uses Boltzmann distributions to select bend angles for spontaneous thermal fluctuations and binding sites for condensing agents. Updating the model only requires updating the interaction energies for these processes. Second, the model produces simulated contours that can be compared to experimental data. Third, the model accounts for both spontaneous looping and looping due to bind and bend. Finally, the model is adjustable. Users can adjust the parameters for the polymer (contour length $L_C$, flexibility $L_P$, and polymer effective charge---$\lambda$ and $p_\text{DNA}$), as well as the parameters for the counterion (effective charge---$q_\text{agent}$ and $p_\text{agent}$, concentration $N_\text{agents}$, and the bend angle threshold $\beta_\text{thresh}$). This allows the application of our model to a number of couterion-polyelectrolyte systems beyond protamine-DNA.

There are also limitations of the model. The model is in 2D rather than 3D, even though bending and looping take place in solution. In addition, the model sets the bend angle so that it is always in the same direction and is set by a threshold value $\beta_\text{thresh}$. Currently, some molecular dynamics simulations suggest that there are different conformations of the protamine-DNA complex \cite{mukherjee2021protamine}. Once the probabilities for these conformations and bend angles are known, the simulation could be updated to model a distribution of this type. Also, the model does not account for DNA-DNA interactions even though the probability of these interactions is likely to be on the same order as the probability for DNA bending \cite{mcmillan_dna_nodate}. Finally, our model does not propagate the bend occurring in one region to other regions based on the persistence length of the polymer.

\subsection*{Implications for DNA looping by protamine}
Importantly, the bind-and-bend model gives insight into how protamine physically loops DNA, answering questions about DNA binding by protamine, the bending of the DNA into a loop, the stability of the DNA-DNA interaction, and the formation of multiple loops.

The first question is about protamine binding. Eventually, protamine will ``coat'' the DNA, but initially, some DNA locations might have higher levels of bound protamine than others, perhaps leading to nucleation of DNA loops or toroids in those locations \cite{hud_toroidal_2005}. Here we see that protamine molecules initially bind to locations where there is a higher DNA concentration (e.g., regions that have high curvature or contain crossover points). However, once a protamine molecule has bound that location, another is not likely to bind, creating an even binding pattern along the DNA. This balance is probably one of the reasons protamine is able to fold the entire sperm genome in minutes \cite{balhorn_protamine_2007, vilfan_formation_2004}. 

The second question is on whether DNA looping by protamine is due to DNA bending or spontaneous thermal fluctuations. Previous models of DNA looping by condensing agents speculated that condensing agents would use spontaneous thermal fluctuations in the DNA to create loops \cite{hud_toroidal_2005}, rather than DNA bending \cite{ukogu2020protamine}. These spontaneous thermal fluctuations are still occurring, but how do they aid looping? Here, we see that spontaneous thermal fluctuations are an integral part of bind-and-bend. Specifically, protamine molecules do not need to bind at every binding site along the circumference of the loop to bend the DNA into a loop. Instead, protamine binds and bends the DNA at some locations within the loop, utilizing thermal fluctuations at other locations within the loop to fold the DNA into loops with a peak circumference of about 200 bp, while also allowing for larger loops.

The third question is on whether a single DNA-DNA interaction that might occur at a loop crossover location is stable. Previous single molecule experiments for condensing agents \cite{mcmillan_dna_nodate, van2010visualizing} have found that DNA folded into a few loops is highly unstable and likely to unfold. Yet, some models predict that a loop is the nucleation event for the DNA toroid \cite{hud_toroidal_2005}, and would presumably have a stable DNA-DNA interaction. Is the DNA loop created by the bind-and-bend mechanism stable? Here we see that folding in 2D requires an unstable DNA-DNA interaction in order for our simulated data sets to agree with the experimental data. It is possible that folding in 3D could alleviate this constraint. Still, even in 3D, unstable DNA-DNA interactions would allow for protamine to create loops with very small radii of $R= 10$ nm, and unstable DNA-DNA interactions might be useful when forming the toroid, as DNA-DNA interactions between loops might have to unform and reform before the loops lock into place. We speculate that stable DNA-DNA interactions (e.g., disulfide bridges between protamines) do not form in the early stages of toroid formation.

Finally, the last question is on whether bind-and-bend leads to the bending of the DNA into multiple loops or flowers. Previous data on DNA looping by condensing agents \cite{mcmillan_dna_nodate, fang1998early} finds the existence of flower structures with multiple loops. It was speculated that the physical mechanism for the formation of these multilooped flowers is DNA bending \cite{mcmillan_dna_nodate}. Here we see that DNA bending does produce flowers, but not at the rate we would expect. We observe only a third of simulated multilooped structures are flowers, while 87\% are reported for the experimental data \cite{mcmillan_dna_nodate}. It may be that we are not capturing all of the physics here. One possibility is that the creation of flowers involves DNA-DNA interactions, which are not accounted for by our model. Another possibility is that twist in the DNA molecule is making flowers more likely. Molecular dynamics simulations show that protamine can bend and twist the DNA \cite{mukherjee2021protamine}, especially if the protamine binds in the minor groove. Future simulations could add in DNA-DNA interactions or twist. 

\subsection*{Broader applications}
Beyond applications to protamine-DNA systems, the bind-and-bend model will likely be generalizable to other condensing agents like cobalt hexaammine (III), spermine, or spermidine \cite{hud_toroidal_2005, schnell1998insertion, bloomfield_condensation_1991, leforestier_structure_2009, leforestier2011protein, takahashi1997discrete, fang1998early, murayama2003elastic, marx1983evidence, takahashi1997discrete} as the theory for DNA bending was first theorized for these condensing agents \cite{rouzina1998dna}. The bind-and-bend model might also be useful for condensation of polyelectrolytes by counterions \cite{muthukumar2004theory, ou2005langevin}, or for use in condensation of hydrogels \cite{lopez2013spermidine, ruseska2021use} or DNA origami nanostructures \cite{fan2017dna}. The model might also be useful for nucleic acid compaction more broadly. HIV-1 nucleocapsid proteins fold DNA into structures that look similar to structures folded by condensing agents \cite{gien2022hiv}.

\section*{Author Contributions}

Michael L. Liu and Ashley R. Carter developed the theory, implemented the model in MATLAB, analyzed the simulated and experimental data, and wrote the article. Daniel W. Oo, Ryan B. McMillan, and Ashley R. Carter collected and analyzed the experimental data.

\section*{Acknowledgements}

This work was supported by a National Science Foundation CAREER award (Project \# 1653501), Clare Boothe Luce, and Amherst College.

\section*{Supporting Material}

Supporting Material can be found online. The accompanying MATLAB code and experimental data are available on Github at \url{https://github.com/MichaelLiu2024}.

\bibliography{Bibliography}

\end{document}


\maketitle

\newpage
\section*{Determining segment intersection points}

\begin{figure}[h]
\includegraphics[width=0.5\linewidth]{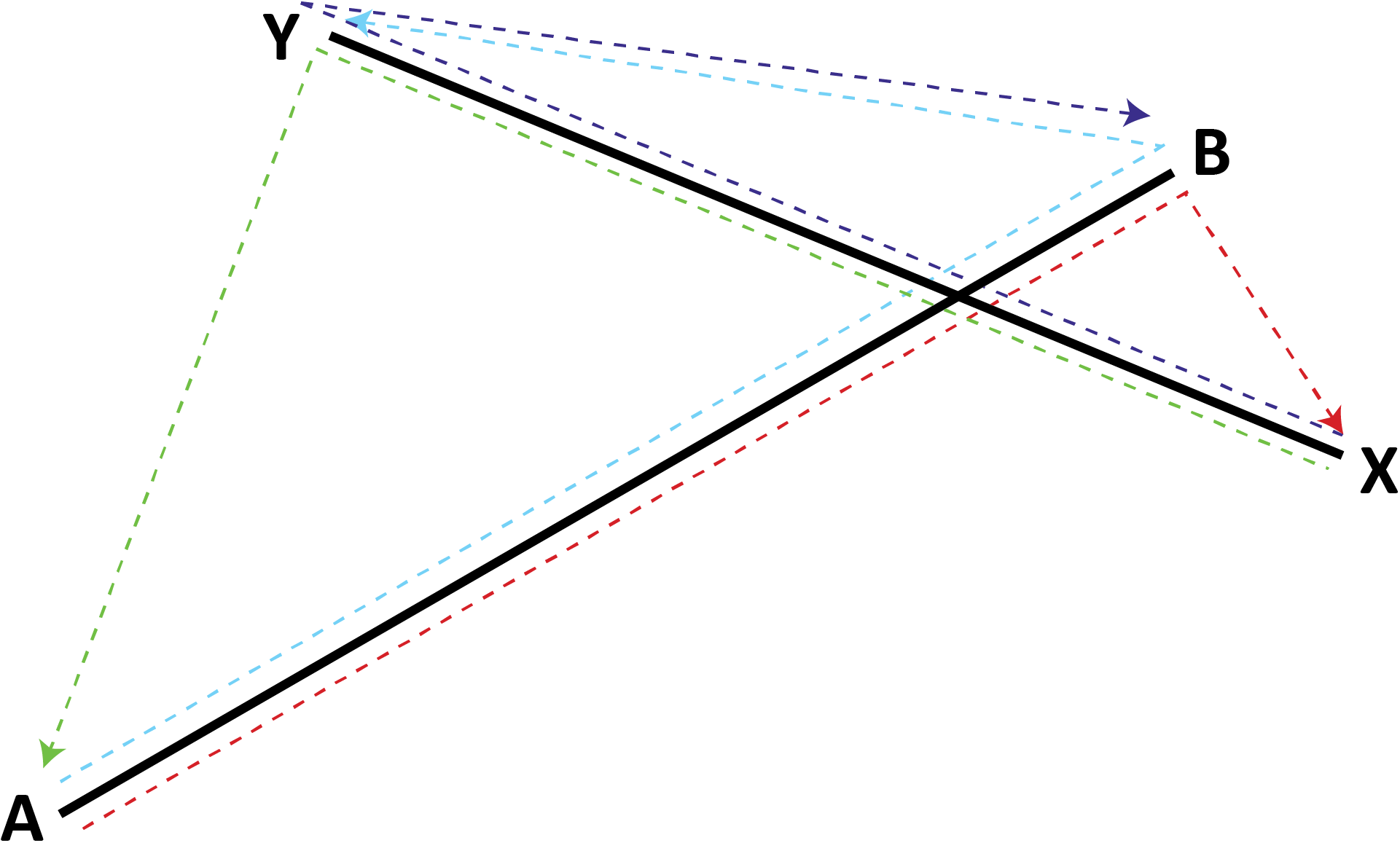}
\caption{\textbf{Algorithm to check for segment intersection.} Two line segments, $AB$ and $XY$, define four angles, $ ABX$, $ ABY$, $ XYA$, and $ XYB$. To determine whether $AB$ and $XY$ intersect, we calculate the orientation of each angle above, either clockwise (CW) or counterclockwise (CCW). Here, $ ABX$ is CW (\emph{red}), $ ABY$ is CCW (\emph{blue}), $ XYA$ is CCW (\emph{green}), and $ XYB$ is CW (\emph{purple}). The two line segments intersect if and only if both pairs ($ ABX$ and $ ABY$, as well as $ XYA$ and $ XYB$) have one CW and one CCW angle.}
\label{fig:segment_intersect}
\end{figure}

Let $AB$ and $XY$ be two line segments in a plane (Fig. \ref{fig:segment_intersect}). We wish to determine whether $AB$ and $XY$ intersect without solving a linear system\cite{cormen2009introduction}. To do this, we first compute the angles $ ABX$, $ ABY$, $ XYA$, and $ XYB$. Next, we observe that $AB$ and $XY$ intersect if and only if the orientations of $ ABX$ and $ ABY$ are different and the orientations of $ XYA$ and $ XYB$ are different. That is, $AB$ and $XY$ intersect if and only if for each pair of angles above, one angle is oriented clockwise, and the other angle is oriented counterclockwise.

Now, consider a discretized representation of the DNA with $n$ binding sites. For all $\genfrac(){0pt}{1}{n+1}{2}$ pairs of line segments in the discretized representation, we check whether that pair of segments intersect using the above algorithm. For those pairs that do intersect, we calculate their intersection point. To calculate the intersection point, we parameterize the two segments. Treating the segment endpoints $\vec A$, $\vec B$, $\vec X$, and $\vec Y$ as vectors, the intersection point is found by solving the two dimensional vector equation
\begin{equation}
    \vec A + s_{AB}(\vec B - \vec A) = \vec X + s_{XY}(\vec Y - \vec X)
\end{equation}
for $s_{AB}, s_{XY} \in [0, 1]$.

\section*{DNA bending by protamine}

Here, our goal is to check whether the DNA bending portion of the simulation fits our experimental data. To complete this task, we recreate Fig. 6 from \cite{ukogu2020protamine}, which has several metrics that allow us to compare our simulated and experimental data set. These metrics include the shape of the DNA contour, the decay of the tangent-tangent correlation, and the mean squared displacement (MSD) for locations along the DNA. To make this comparison, we simulated 100 molecules of DNA with contour length $L_C= 309$ bp with and without protamine (parameters in Rows 3-4 of Table \ref{tab:parameters}). Then, we compared these simulated molecules to our AFM images of DNA molecules ($L_C= 309$ bp). These molecules are immobilized on the surface with magnesium acetate in the presence (0.2 $\mu$M) and absence of protamine. To compute the contours, tangent-tangent correlation, and MSD for both data sets we used the ``combo\_calc'' function in the Easyworm software suite \cite{lamour2014easyworm}.

In the first metric (Fig. \ref{fig:protbending_all}A), we plot contours of the DNA ($L_C= 309$ bp). We align the contours so that one end of the DNA starts at the origin, and so that the DNA points along the $x$ axis. In both the experimental and simulated data without protamine (\emph{gray}), the molecules appear fairly straight, continuing in the direction of $x$ and often not deviating from the $x$ axis by more than $45^{\circ}$. In comparison, the shape of the DNA contour for both the experimental and simulated data in the presence of protamine (\emph{yellow}) appears as a ``C" shape, with the molecules folding over onto themselves, often folding past $90^{\circ}$. We do see one difference: folding in the simulation creates molecules that are always bent in the same direction (always CW or always CCW), whereas folding in the experimental data shows some bending in both directions (both CW and CCW in the same molecule). This aspect of the simulation is by design.

The second metric is computing the tangent-tangent correlations. To compute the tangent-tangent correlation, we look at two locations along the DNA with an arc distance of $s$ between the them. We then compute the tangent at each location and measure the angle $\theta$ between the two tangents. If the DNA is a worm-like chain, then for small $s$ (less than the persistence length $L_P$ of the DNA), the cosine of $\theta$ should be about one and the tangents are correlated. As we increase $s$, the two tangents start to become uncorrelated with angles anywhere between $0^{\circ}$ and $180^{\circ}$. The average angle is $90^{\circ}$ and the average cosine of the angle is zero. Thus, for a length of DNA that is well modeled as a worm-like chain, the expectation value of the cosine of $\theta$ should decay exponentially from one to zero as the distance $s$ along the DNA contour increases. Explicitly, we have
\begin{equation} \label{eq:CosThetaWLC}
\langle\cos(\theta)\rangle = e^{-\frac{s}{L_P}}, 
\end{equation}
where the decay factor $L_P$ is the persistence length of the DNA. On the other hand, if the DNA is being bent by protamine into an arc, then the tangent around the arc follows the prescribed pattern of a circle. The $\theta$ for two adjacent DNA locations on the circle should be approximately zero. Then, $\theta$ changes to $90^{\circ}$ as the locations become separated by a quarter of the circumference, and it changes to $180^{\circ}$ at half the circumference. This means that the expectation for the cosine of $\theta$ for this data should be given by a circular arc,
\begin{equation} \label{eq:CosThetaCircle}
\langle\cos(\theta)\rangle = \cos \left(\frac{s}{R} \right),
\end{equation}
where $R$ is the radius of the circle. 

In both the simulated and experimental data sets (\ref{fig:protbending_all}B), we see that in the absence of protamine (\emph{gray}), the DNA follows Eq. \ref{eq:CosThetaWLC} (\emph{solid black}) with an $L_P$ of 60 nm, indicating it can be approximated as a worm-like chain. However, when we add protamine (\emph{yellow}) both the experimental and simulated distribution decay rapidly, crossing zero so that the tangents are negatively correlated. A fit of Eq. \ref{eq:CosThetaWLC} (\emph{solid brown}) with a very low $L_P$ of 10 nm does not fit either curve. However, the equation for DNA being bent by protamine into a perfectly circular arc as in Eq. \ref{eq:CosThetaCircle} (\emph{dashed brown}) does fit the data. The fit captures both the crossing of the $x$ axis and the change from concave down to concave up at about that point. 

Interestingly, we see that the fit is better in the simulation since more molecules cross the $x$ axis. This is probably because there is some bending in the opposite direction in the experimental data, but not in the simulated data. 

The third metric we consider is the MSD between points on the DNA contour. For a worm-like chain, the conformation of the DNA should essentially follow a random walk with a step size related to the persistence length $L_P$ of the DNA. Thus, the MSD should increase as the length along the DNA contour $s$ between the two DNA locations increases. Specifically, the MSD should be a function of $s$ and $L_P$, such that
\begin{equation} \label{eq:MSDWLC}
\text{MSD}(s) = 2 L_P \times s \left(1 - \frac{L_P}{s} \left(1-e^{-\frac{s}{L_P}}\right) \right).
\end{equation}

\noindent However, the MSD for a circle should increase to a maximum when $s$ is half the circumference, and then decrease back to zero as $s$ approaches the circumference of the loop. The MSD in this case is
\begin{equation} \label{eq:MSDCircle}
\text{MSD}(s) = 2 R^2 \left (1 - \cos \left(\frac{s}{R} \right) \right).
\end{equation}

We compute the MSD for the simulated and experimental DNA molecules, and check the fits against both Eq. \ref{eq:MSDWLC} and Eq. \ref{eq:MSDCircle} (Fig. \ref{fig:protbending_all}).  We expect the worm-like chain model, Eq. \ref{eq:MSDWLC} (\emph{solid black line}), to hold for DNA molecules in the absence of protamine (\emph{gray}), and it does in both the experimental and simulated data sets. Then, in the presence of protamine (\emph{yellow}), we observe that the DNA is bent into a circle as in Eq. \ref{eq:MSDCircle} (\emph{dashed brown line}).

For all three metrics, our simulated and experimental data sets agree. We therefore conclude that our bind-and-bend model is capturing protamine bending.

\clearpage
\section*{Figures}

\begin{figure}[h]
\includegraphics[width=0.80\linewidth]{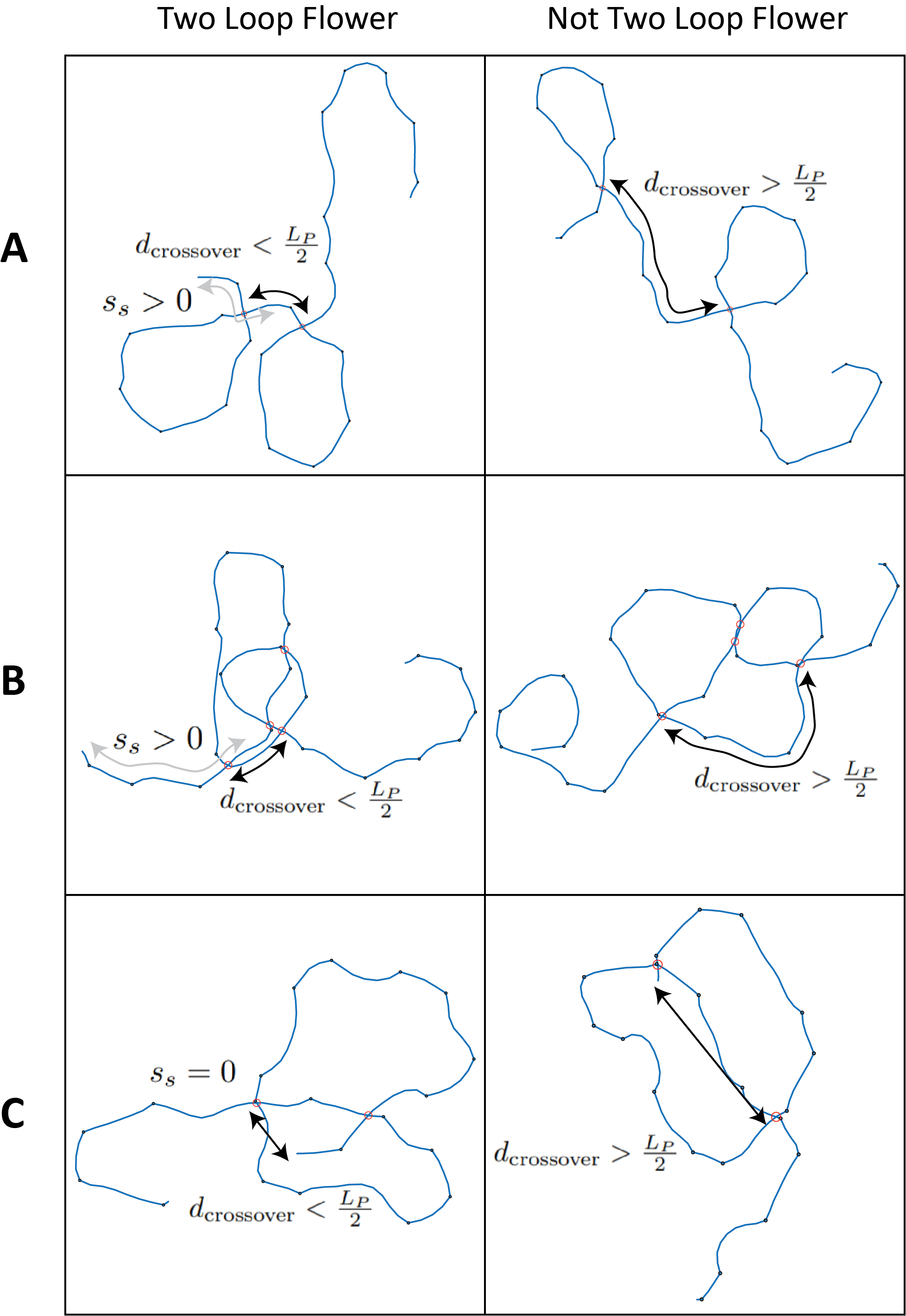}
\caption{\textbf{Examples of Two Loop Flowers.} A) Separate loops form a flower if their crossover points are within $\frac{L_P}{2}$ of each other along the DNA contour. Black arrows show the crossover point separation; grey arrows show the start site measurement. B) Overlapping loops also form a flower given their crossover points are within $\frac{L_P}{2}$ of each other. C) Flowers can form by thermal loop closing if a DNA endpoint is within $\frac{L_P}{2}$ of a loop crossover point.}
\label{fig:two_loops}
\end{figure}

\begin{figure}[h]
\includegraphics[width=0.80\linewidth]{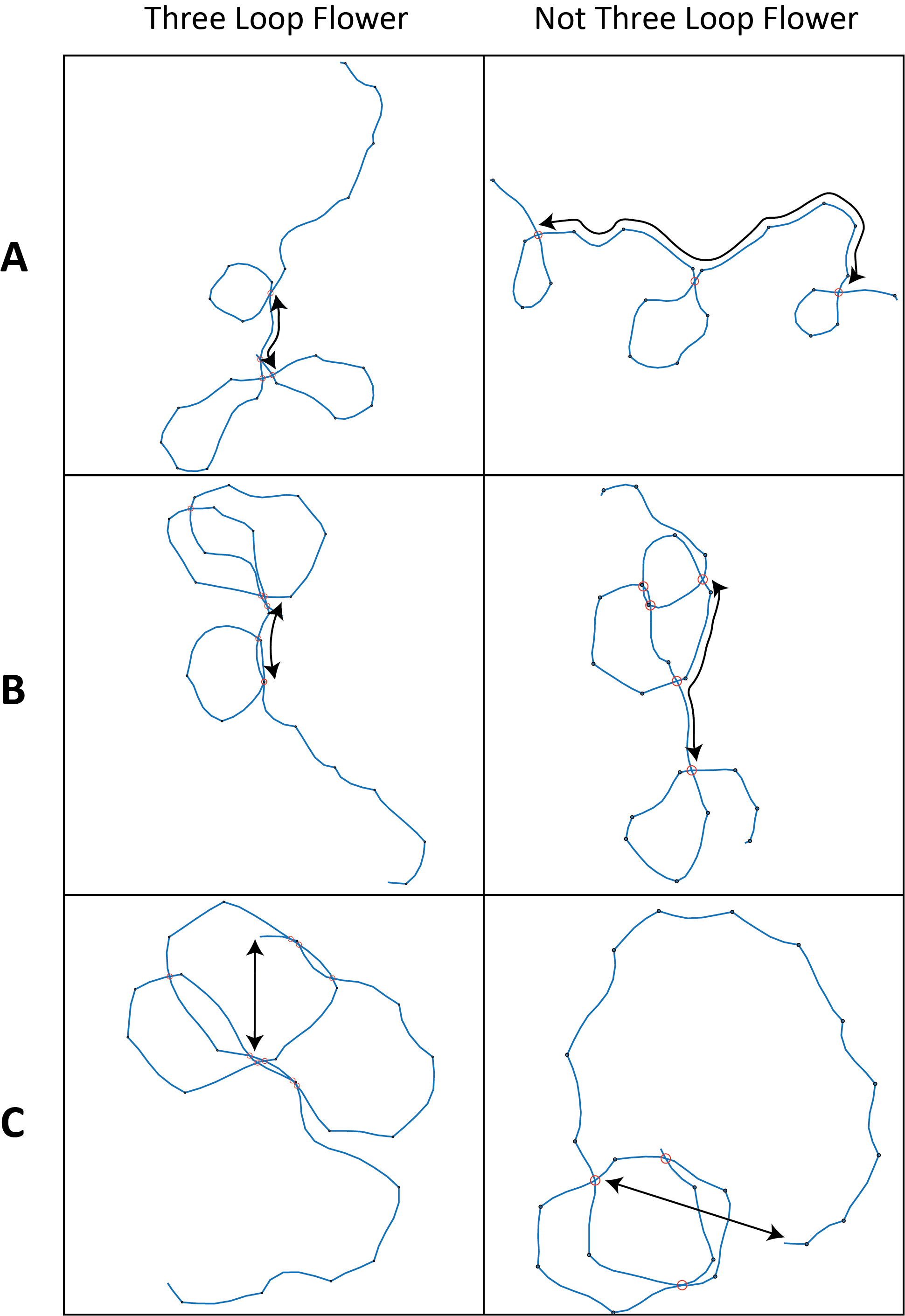}\caption{\textbf{Examples of Three Loop Flowers.} A) In a molecule with three distinct loops, the most extreme loop crossover points must be within $\frac{L_P}{2}$ of each other for the molecule to be counted as a three loop flower. B) Molecules with three overlapping loops can also form a flower if their most extreme crossover points are within $\frac{L_P}{2}$ of each other. C) Three loop flowers can form by thermal loop closing if a DNA endpoint lies within $\frac{L_P}{2}$ of a crossover point.}
\label{fig:three_loops}
\end{figure}

\begin{figure}[h]
\includegraphics[width=0.80\linewidth]{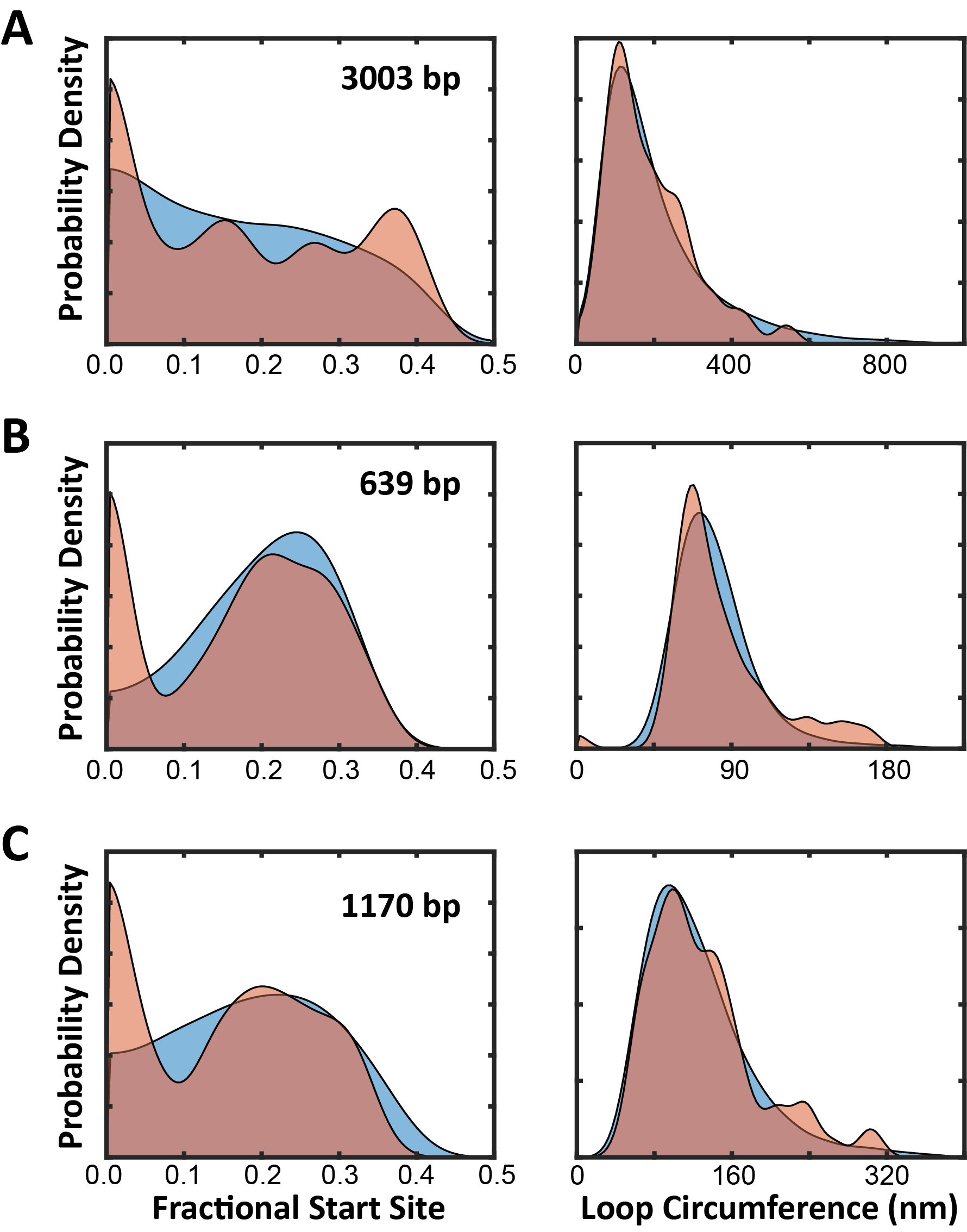}
\caption{\textbf{Fits without thermal loop closing.} Simulated (\emph{blue}) and experimental (\emph{red}) distributions of the fractional distance to the start site $s_s/L_C$ (\emph{left}) and the loop circumference $c$ (\emph{right}) for (A) spontaneously looped 3003 bp DNA, (B) protamine-induced looping of 639 bp DNA, and (C) protamine-induced looping of 1170 bp DNA, all without thermal loop closing. In all cases, when thermal loop closing is not present, the distribution of the fractional distance to the start site loses the peak at $s_s/L_C = 0$. }
\label{fig:no_loopclosing}
\end{figure}

\begin{figure}[h]
\centering
\includegraphics[width=0.60\linewidth]{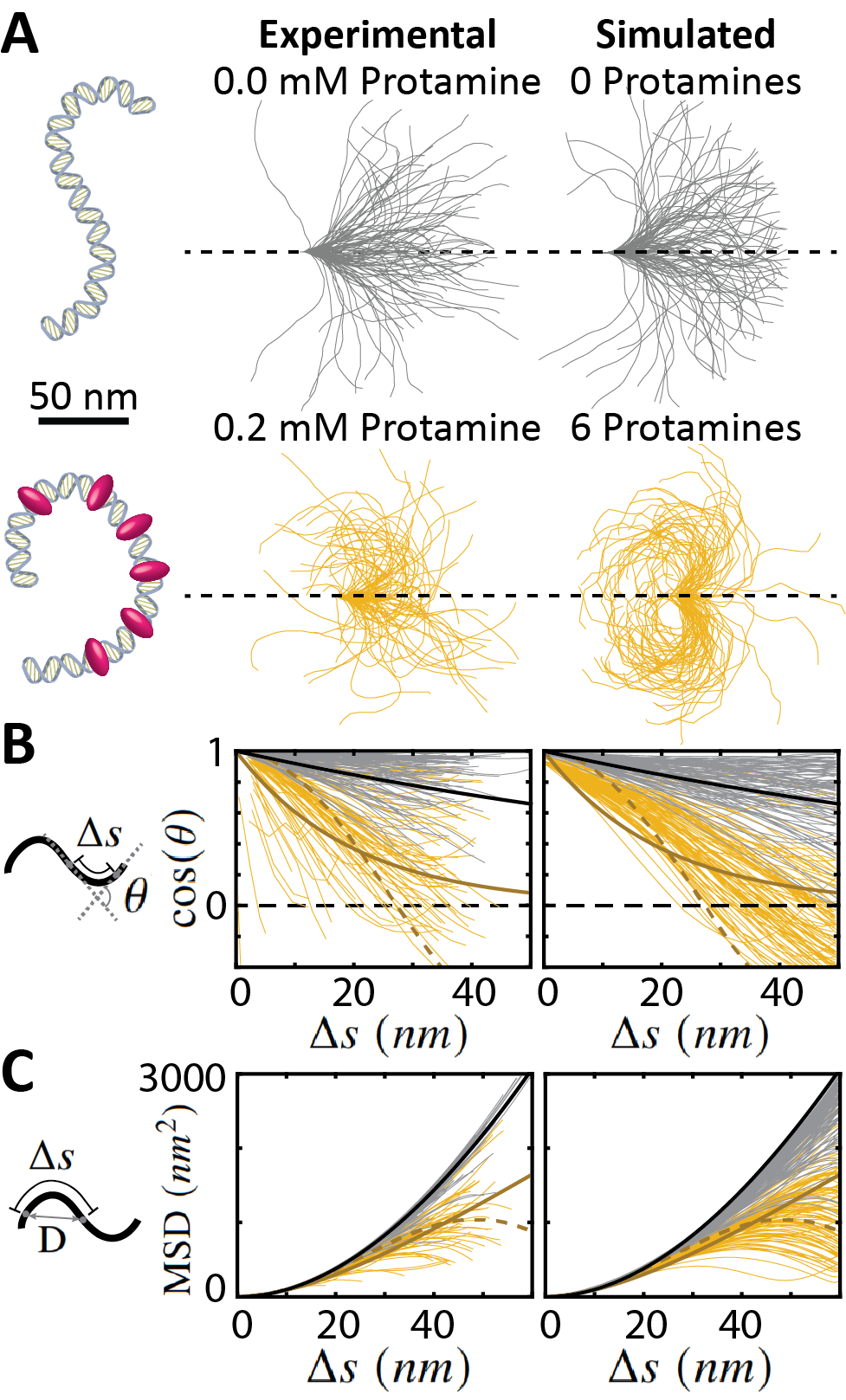}
\caption{\textbf{Model accurately simulates protamine-induced DNA bending.} We compare experimental and simulated data for 309-bp-long DNA molecules with protamine (yellow) and without protamine (grey). To make the comparison, we repeat previous experimental measurements \cite{ukogu2020protamine} on our simulated DNA contours. A) DNA contours from AFM images ($N_\text{0.0{\textmu}M} = 95$, $N_\text{0.2{\textmu}M} = 93$) and from the simulation ($N_0 = N_6 = 100$) are aligned so that they start parallel to the $x$ axis. Molecules with protamine tend to loop back toward the origin. B) Plot of the tangent-tangent correlation $\cos(\theta)$ against the displacement $\Delta s$ along the DNA contour. Fits according to Eq. \ref{eq:CosThetaWLC} (bold lines) tend toward zero, while fits according to Eq. \ref{eq:CosThetaCircle} (dashed lines) cross the $x$ axis. C) Plot of the mean squared displacement against $\Delta s$. Fits according to Eq. \ref{eq:MSDWLC} (bold lines) grow monotonically, while fits according to Eq. \ref{eq:MSDCircle} (dashed lines) reach a maximum at $\Delta s \approx 50nm$. In both panels B and C, the data with protamine follows Eq. \ref{eq:CosThetaCircle} and Eq.~\ref{eq:MSDCircle}, which indicates that the DNA is being bent into a circle with a radius of curvature of 10-20 nm.}
\label{fig:protbending_all}
\end{figure}

\begin{figure}[!htb]
\centering
\includegraphics[width=0.80\linewidth]{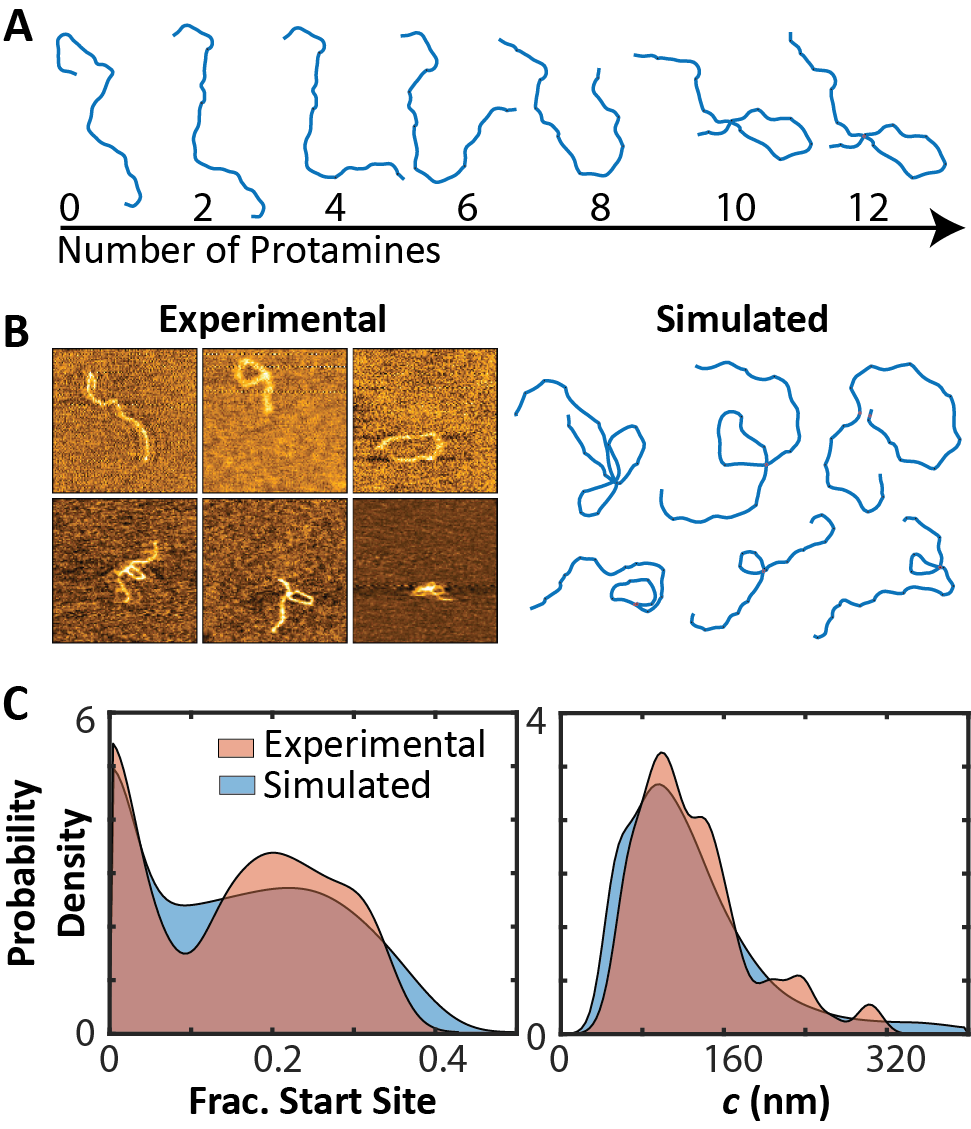}
\caption{\textbf{Model accurately simulates protamine-induced loop formation.} We simulate 1170-bp DNA molecules using Row 7 of Table \ref{tab:parameters}.  A) An unlooped 1170-bp DNA molecule is folded by the addition of protamine molecules to form a single loop. B) Experimental data from AFM images of 1170-bp DNA in the presence of 0.2 {\textmu}M protamine and simulated contours for the same DNA length with $N_\text{agents}= 15$ molecules. C) The experimental (red) and simulated (blue) distributions of fractional start site distance and loop circumference $c$.}
\label{fig:SingleLoops1170bp}
\end{figure}

\begin{figure}[h]
\includegraphics[width=\linewidth]{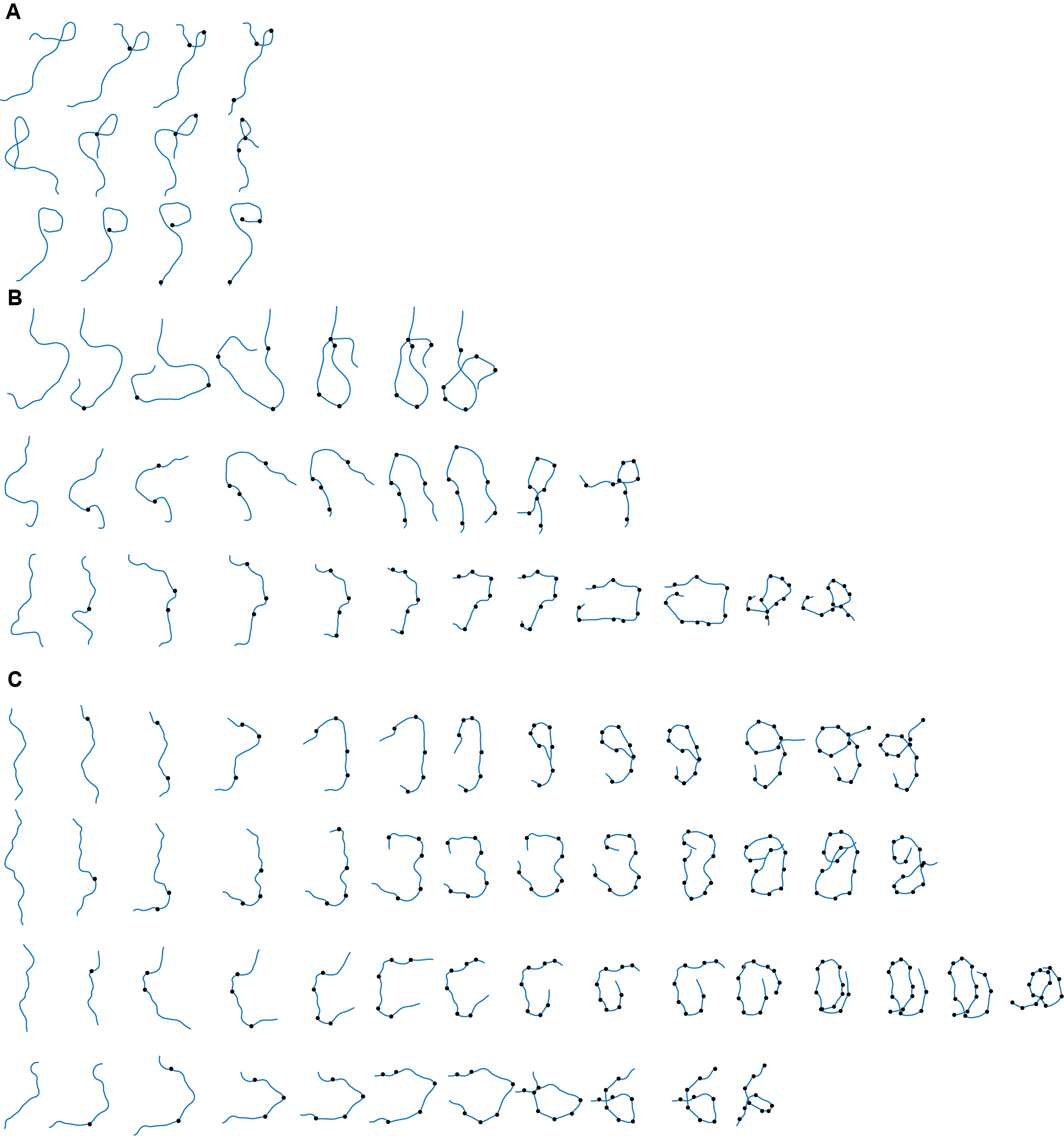}
\caption{\textbf{Representative, simulated DNA contours show how protamine binds and bends DNA.} DNA contours (639-bp-long) from the simulation using parameters in Row 5 of Table \ref{tab:parameters}. A) DNA molecules (\emph{blue curves}) that start with a loop formed by spontaneous thermal fluctuations. The first protamine (\emph{black dot}) that binds this type of DNA molecule usually binds near the DNA crossover. Subsequent protamine molecules will bind in the loop since there is a high DNA concentration there. However, if the loop is small (less than 70 nm circumference or so), then protamine is not likely to bind there again.  B) DNA molecules that start with some folded regions. The first protamine that binds this type of DNA molecule usually binds to a location with a high curvature. Subsequent protamines spread out along the DNA and begin bending the DNA into a loop. After a loop forms, more protamine will bind within the loop, bending it into a smaller loop. C) DNA molecules that start out fairly straight. Protamine binds this type of DNA molecule so that the protamine spreads out along the DNA. After some protamine binding events, the DNA eventually folds onto itself creating a large loop. Subsequent protamines continue to bend the DNA, creating a smaller loop. Contours not to scale. }
\label{fig:contours}
\end{figure}

\begin{figure}[h]
\centering
\includegraphics[width=0.80\linewidth]{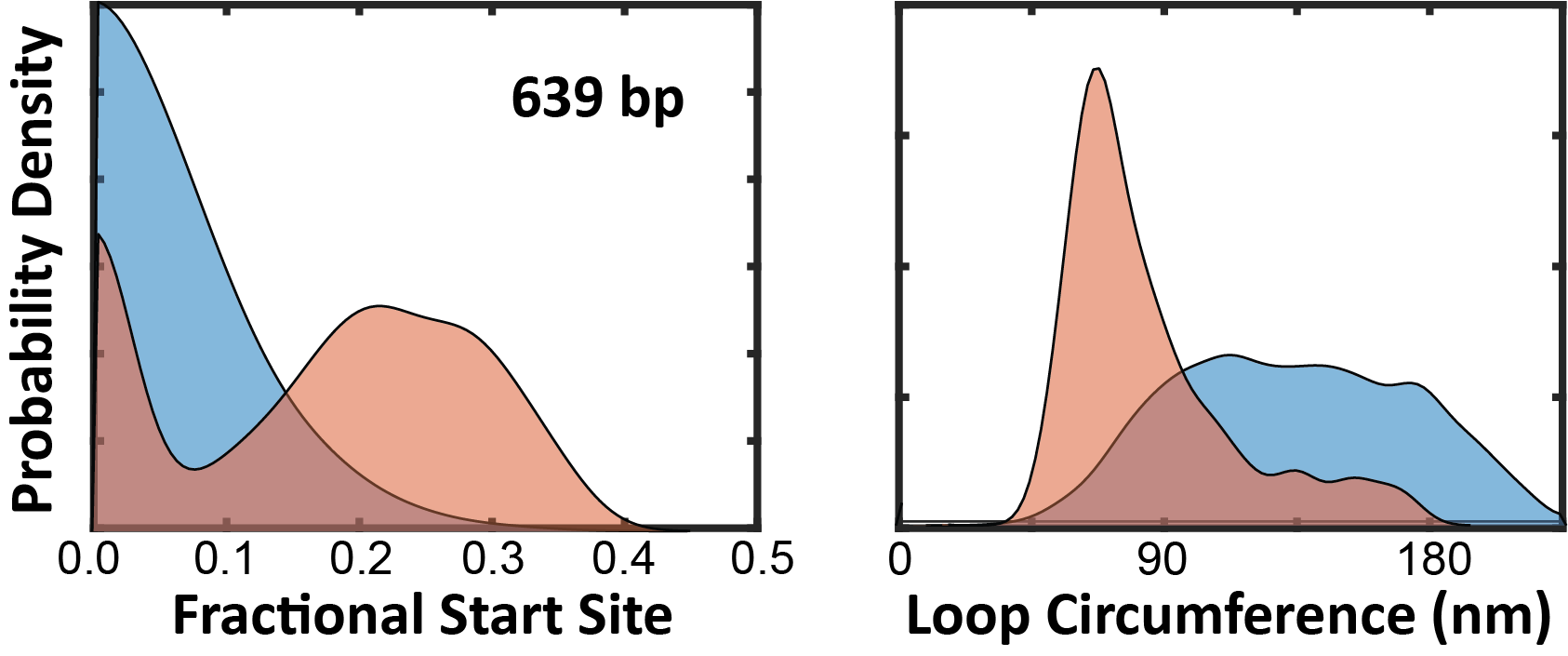}
\caption{\textbf{Properties of loops when they first form.} Simulated (\emph{blue}) and experimental (\emph{red}) distributions of the fractional distance to the start site $s_s/L_C$ (\emph{left}) and the loop circumference $c$ (\emph{right}) for protamine-induced looping of 639-bp DNA. In these plots, we stop the simulation of each DNA contour as soon as a loop forms. These newly formed loops have a larger loop circumference and a smaller fractional start site on average than the loops that form after many ($>$10) protamine molecules have bound.}
\label{fig:StopAtLoop}
\end{figure}

\clearpage
\section*{Tables}

\begin{table}[h]
\centering
\begin{tabular}{ccccccccc}
\hline
Row & Figure & $L_C \ (bp)$ & $L_C \ (nm)$ & $L_P \ (nm)$ & $k_\text{planar}$ & $k_\text{linear}$ & $N_\text{agents}$ & $N_\text{trials}$ \\ \hline
1 & Fig. 2 & $3003$ & $1023$ & $50$ & $0.5$ & $2$ & $0$ & $100,000$ \\
2 & Fig. \ref{fig:no_loopclosing}A & $3003$ & $1023$ & $50$ & $0$ & $2$ & $0$ & $100,000$ \\
3 & Fig. 3 & $309$ & $105$ & $120$ & N/A & N/A & $0$ & $100$ \\
4 & Fig. 3 & $309$ & $105$ & $120$ & N/A & N/A & $6$ & $100$ \\
5 & Fig. 4 & $639$ & $217$ & $50$ & $1$ & $1$ & $18$ & $100,000$ \\
6 & Fig. \ref{fig:no_loopclosing}B & $639$ & $217$ & $50$ & $0$ & $1$ & $18$ & $100,000$ \\
7 & Fig. \ref{fig:SingleLoops1170bp} & $1170$ & $398$ & $50$ & $1$ & $1$ & $15$ & $100,000$ \\
8 & Fig. \ref{fig:no_loopclosing}C & $1170$ & $398$ & $50$ & $0$ & $1$ & $15$ & $100,000$ \\
9 & Fig. 5& $1170$ & $398$ & $50$ & 1 & 1 & $20$ & $2,000$ \\
10 & Fig. 5 & $1170$ & $398$ & $50$ & 1 & 1 & $25$ & $2,000$ \\
 \hline
\end{tabular}
\caption{\textbf{Simulation parameters for the bind-and-bend model.} This table lists the parameter values for simulations that produce data for the figure and DNA length referenced. Only parameters appearing in this table are changed across different simulations. ``N/A" values appear for the threshold factors $k_\text{planar}$ and $k_\text{linear}$ if thermal loop closing was not used to identify loops. }
\label{tab:parameters}
\end{table}

\begin{table}[h]
\centering
\begin{tabular}{ccccccc}
\hline
$L \ (nm)$ & $q_\text{agent} \ (e)$ & $\lambda \ (e/nm)$ & $p_\text{agent}$ & $p_\text{DNA}$ & $p_\text{site}$ & $\beta_\text{thresh} (^\circ)$ \\ \hline
$3.4$ & $21$ & $5.88$ & $0.2$ & $0.01$ & $0.05$ & $47$ \\ \hline
\end{tabular}
\caption{\textbf{Constant values for the bind-and-bend model.} This table lists constant values used in the simulation. The unit of elementary charge is denoted $e$.}
\label{tab:constants}
\end{table}

\newpage
\printbibliography